\documentclass[5p,twocolumn,times]{elsarticle} 

\usepackage{siunitx}
\usepackage{makecell}
\usepackage{xcolor}
\usepackage{hyperref}
\usepackage[ruled,vlined]{algorithm2e}
\usepackage{ragged2e, microtype}
\usepackage{subcaption}
\usepackage{pifont}
\usepackage{amssymb}
\usepackage{float}

\begin{document}

\newcommand{\yestick}{{\color{olive}\ding{51}}}
\newcommand{\notick}{{\color{red}\ding{55}}}

\title{\LARGE \bf
A methodology to identify identical single-board computers based on hardware behavior fingerprinting}

\author[1]{Pedro Miguel {S\'anchez S\'anchez}\corref{cor1}%
}

\author[1]{Jos\'e Mar\'ia {Jorquera Valero}}

\author[2]{Alberto {Huertas Celdr\'an}}

\author[3]{G\'er\^ome Bovet}

\author[1]{Manuel {Gil P\'erez}}

\author[1]{Gregorio {Mart\'inez P\'erez}}

\address[1]{Department of Information and Communications Engineering, University of Murcia, Murcia 30100, Spain}

\address[2]{Communication Systems Group (CSG), Department of Informatics (IfI), University of Zurich UZH, 8050 Zürich, Switzerland}

\address[3]{Cyber-Defence Campus, armasuisse Science \& Technology, 3602 Thun, Switzerland}

\cortext[cor1]{Corresponding author.
Email address: pedromiguel.sanchez@um.es (P.M.S. S\'anchez)}

\begin{abstract}

The connectivity and resource-constrained nature of single-board devices open the door to cybersecurity concerns affecting Internet of Things (IoT) scenarios. One of the most important issues is the presence of unauthorized IoT devices that want to impersonate legitimate ones by using identical hardware and software specifications. This situation can provoke sensitive information leakages, data poisoning, or privilege escalation in IoT scenarios. Combining behavioral fingerprinting and Machine/Deep Learning (ML/DL) techniques is a promising approach to identify these malicious spoofing devices by detecting minor performance differences generated by imperfections in manufacturing. However, existing solutions are not suitable for single-board devices since they do not consider their hardware and software limitations, underestimate critical aspects such as fingerprint stability or context changes, and do not explore the potential of ML/DL techniques. To improve it, this work first identifies the essential properties for single-board device identification: uniqueness, stability, diversity, scalability, efficiency, robustness, and security. Then, a novel methodology relies on behavioral fingerprinting to identify identical single-board devices and meet the previous properties. The methodology leverages the different built-in components of the system and ML/DL techniques, comparing the device internal behavior with each other to detect variations that occurred in manufacturing processes. The methodology validation has been performed in a real environment composed of 15 identical Raspberry Pi 4 Model B and 10 Raspberry Pi 3 Model B+ devices, obtaining a 91.9\% average TPR with an XGBoost model and achieving the identification for all devices by setting a 50\% threshold in the evaluation process. Finally, a discussion compares the proposed solution with related work, highlighting the fingerprint properties not met, and provides important lessons learned and limitations.
\end{abstract}


\maketitle

\section{INTRODUCTION}
\label{sec:intro}

The diversity of IoT devices in modern scenarios is huge, but single-board devices, such as Raspberry Pi, have gained enormous prominence due to their flexibility, reduced price, broad support, and peripherals availability \cite{fayos2020performance}. Unfortunately, the connectivity and resource-constrained nature of single-board devices, and IoT in general, open the door to numerous cybersecurity concerns affecting heterogeneous platforms \cite{gomez2019generation}. One of the most important cybersecurity concerns affecting IoT is the presence of unauthorized devices with the same hardware and software configuration as authorized nodes. Some real attacks based on unauthorized devices have caused big impacts in areas such as Industry 4.0 \cite{jagdale2022new} or mobile phones \cite{Montalbano2022new}. These malicious devices can be articulated by several well-known cybersecurity threats \cite{liu2020zero}, such as \textit{device spoofing} \cite{nosouhi2022towards}, occurring when an attacker replaces a legitimate sensor or actuator with a malicious device using the same identity; \textit{unauthorized device deployment}, related to the installation of a new device in the platform which is using an unregistered identity; and \textit{Sybil attack}, referring to a malicious device (or many) using numerous identities to simulate being several devices. After that, other cybersecurity threats, such as \textit{sensitive information leakage}, \textit{data poisoning}, or \textit{privilege escalation and lateral movements}, might arise as a consequence of spoofed devices. Besides, modern attacks exploit evasion techniques in order to be undetected by software-based security methods.

Traditional identification solutions rely on names, identifiers, certificates, or tags in order to perform the identification tasks. However, these solutions can be cloned or modified if the software of the device is completely replicated \cite{yousefnezhad2020security}.  Hardware behavior fingerprinting is a potential solution for the identification of single-board devices with identical hardware and software, but still an emergent and open challenge. In such a context, there is no work focused on identical single-board device identification \cite{sabhanayagam2022comparative}. However, for other devices without component and resource limitations, the literature has proposed the usage of hardware \textit{behavioral fingerprinting} as a promising solution to detect minor performance differences generated by imperfections occurred during the devices manufacturing process \cite{al2018survey}.

In particular, existing work focuses on crystal oscillator impurities and cut variations that generate imperfect frequency outputs in components such as CPU or GPU to detect performance differences in identical devices \cite{polcak2015clock}. Current solutions consider dimensions such as clock-skew analysis, intrinsic Physical Unclonable Functions (PUFs), or execution time and performance analysis \cite{Sanchez2018clock}. However, despite their benefits, the following challenges are still open: i) many solutions require additional components or modifications in the devices, which is not possible in some IoT scenarios ii) there is no solution for identifying identical single-board devices based on their hardware; iii) existing solutions are designed for traditional computers, being not suitable for IoT environments with single-board devices with software and hardware restrictions; iv) most of the existing identification solutions have been tested missing essential properties and requirements affecting the identification performance; and v) despite Machine and Deep Learning (ML/DL) techniques have gained enormous importance for the last years, they have not yet been widely applied in the individual device identification field \cite{sanchez2020survey}.

In order to improve the previous challenges, the main contributions of the present work are:
\begin{itemize}
    \item The definition of a set of properties that should be fulfilled by any fingerprinting solution in charge of identifying identical single-board devices. These properties are uniqueness, stability, diversity, scalability, efficiency, robustness, and security.

    \item A novel methodology that leverages hardware behavioral fingerprinting to identify identical single-board devices, solving the problems and drawbacks of previous solutions. Some of these problems are the need for additional hardware, chip modifications, or physical access to the device to perform the identification. The proposed methodology creates unique device behavioral fingerprints measuring the impact that insignificant hardware differences, happened during the manufacturing process of identical devices, have on the device performance when a given task is executed. 
    
    \item The validation of the proposed methodology, as a Proof-of-Concept (PoC) available on \cite{code}, in a scenario composed of 25 identical Raspberry Pi 3 and 4 devices used in IoT scenarios. After testing different ML/DL algorithms, 91.9\% average TPR was achieved by XGBoost, and a perfect identification was carried out by setting a 50\% threshold in the assigned classes. 
    
    \item A detailed analysis and comparative of existing device fingerprinting solutions for individual device identification, focusing on their suitability for IoT environments with single-board devices. It highlights which fingerprint properties are not met in each solution, rising issues such as reproducibility and solution stability.
\end{itemize}

The remainder of this paper is organized as follows. Section \ref{sec:related} reviews the main solutions for identical device identification and discusses why these approaches are not appropriate for IoT environments based on single-board devices. Section \ref{sec:problem} describes the problem to be solved by the present methodology. It details a short threat model for the single-board device identification scenario. Then, it details the set of properties required in a fingerprinting solution to make it appropriate for individual device identification, together with the limitations found in the literature works. The design of the proposed device identification methodology is explained in Section \ref{sec:methodology}, verifying how each fingerprint property is accomplished. Section \ref{sec:PoC} acts as validation of the present methodology, implementing it as a PoC that verifies its applicability in a realistic use case. Section \ref{sec:discussion} compares the literature works with the proposed methodology and depicts several lessons learned and limitations. Finally, Section \ref{sec:conclusions} shows the conclusions extracted from the present work and future steps in the research.

\section{RELATED WORK}
\label{sec:related}

This section gives the main insights of the related work dealing with unique device identification, paying special attention to device identification without additional external hardware requirements.

\begin{table*}[htpb]
    \centering
    \scriptsize
    \caption{Individual device identification solutions based on device behavior fingerprinting.}
    \label{tab:identical_device_table}
    \begin{tabular}{ >{\Centering}m{0.6cm} >{\Centering}m{0.6cm} >{\Centering}m{1.4cm} >{\Centering}m{1.7cm} >{\Centering}m{1.7cm} >{\Centering}m{2.3cm} >{\RaggedRight\arraybackslash}m{6.8cm} } 
    \hline
    \textbf{Work} & \textbf{Year} & \textbf{Device Type} & \textbf{Algorithms} & \textbf{Behavior Source} & \textbf{Features} & \makecell[c]{\textbf{Results}} \\
    \hline
    \cite{salo2007multi} & 2007 & General computers &  Statistical & Processors and oscillators & RTC and DSP drift compared to the TSC & 98.5\% and 93.3\% of differentiation by RTC and DSP in 38 PCs, respectively. \\
    \hline
    \cite{jana2009skew} & 2009 & Wireless access points &  Expectation Maximization & Clock skew & Wi-Fi beacons timestamps & Clock skew is a robust method and can detect different WLAN APs. \\
    \hline
    \cite{sharma2012skew} & 2012 & General computers &  Statistical & Clock skew & TCP and ICMP timestamp & Both identical and different devices correctly identified. \\ 
    \hline
    \cite{wang2012flash} & 2012 & General computers & Correlation coefficient & Flash memory & Bit partial programming & Estimated false positive chance of \num{4.52e-815}, and a false negative chance of \num{2.65e-539}.\\
    \hline
    \cite{radhakrishnan2014gtid} & 2014 & Wireless devices &  ANN & Clock skew + Network & Communi-cation skew and patterns & From 99 to 95\% accuracy and 74\% recall on individual classification. \\
    \hline
    \cite{nakibly2015hardware} & 2015 & General computers & Entropy & GPU & Frames per second & Graphic rendering show differentiation capabilities on 9 identical PCs, but no advanced tests were performed. \\
    \hline
    \cite{Sanchez2018clock} & 2018 & General computers &  Statistical (Mode) & System processors & Matrix of code execution times & 100\% host-based and +80\% web-based device identification in two sets of 89 and 176 PCs. \\
    \hline
    \cite{Jafari_Fingerprinting_Deep_Learning_2018} & 2018 & Wireless devices &  MLP, CNN, LSTM & Electromagnetic signals &  Radio frequency IQ samples & 96.3\% accuracy for MLP, 94.7\% for CNN and 75\% for LSTM when identifying 6 identical ZigBee devices.\\
    \hline
    \cite{Riyaz2018Radio} & 2018 & Wireless devices &  CNN & Electromagnetic signals & Raw frequency IQ samples & 98\% accuracy is achieved when identifying 5 identical devices. \\
    \hline
    \cite{dong2019cpg} & 2019 & General computers &  Dynamic Time Warping & Resource usage & CPU usage- based graph & 93.43\% of uniqueness in the generated fingerprints of 10 identical devices. \\
    \hline
    \cite{BabunCPS2021} & 2021 & CPSs & Correlation-based (Own) &  Hardware and OS/kernel & Syscalls, Memory, CPU, Time & Device type (model/OS version) identification, not individual identification.\\
    \hline
    This Work & 2022 & Single-board devices & XGBoost & Hardware cycle counter skew & Window-based GPU/CPU features & 91.9\% average TPR when identifying 15 RPi4 and 10 RPi3 devices. \\
    \hline
    \end{tabular}
\end{table*}

As a main remark, it is worth mentioning that, to the best of our knowledge, there is no methodology for individual fingerprinting of IoT devices based on hardware characteristics. In fact, the same happens in the field of traditional devices such as personal computers. In this regard, the closest work is the one proposed by Babun et al. \cite{BabunCPS2021}, in which a fingerprinting framework for identifying classes of Cyber-Physical Systems (CPSs) was presented. This solution employed hardware and OS/kernel characteristics following a challenge/response mechanism for performance and system calls fingerprinting. During the validation, a set of single-board computers were employed. Nevertheless, the objective of this framework is device type (class) fingerprinting and identification, not individual device fingerprinting when hardware and software are identical. Therefore, following this approach, identical devices would generate the same fingerprints, as the data sources leveraged are based on OS/kernel or component-related data and do not seek to identify fabrication variations or imperfections.

Although not defined in the form of a methodology, it is essential to analyze existing work focused on individual device fingerprinting and identification for other types of devices, discussing why it is not appropriate for single-board devices. In this context, traditionally, Physical Unclonable Functions (PUFs) have been one of the main methods for unique device identification. PUFs are hardware elements that generate a unique physically-defined fingerprint for a given output based on the manufacturing characteristics of the physical chips. PUFs have been employed in IoT from several perspectives \cite{babaei2019physical}. However, PUFs require additional dedicated hardware elements that have to be attached to the device, making this solution not scalable in large environments or where direct access to the device is not possible. In contrast, intrinsic PUFs in the literature require hardware modifications \cite{kong2013processor} or components such as SRAM not present in IoT devices due to cost restrictions \cite{gao2019building}.

From crystal oscillator analysis, Salo \cite{salo2007multi} exploited differences in Real-Time Clocks (RTCs) and sound card Digital Signal Processors (DSPs) based on the drift between these chips and the CPU cycle counter (TSC in Intel processors). RTC-based and DSP-based differentiation achieved 98.7\% and 93.3\% of uniqueness when 703 computer pairs were evaluated. However, this method involves the use of components that, although common in computers, are not often available in single-board devices. Also leveraging oscillators, Sanchez-Rola et al. \cite{Sanchez2018clock} proposed a fingerprinting method based on execution time. The authors cyclically executed a simple function to generate a time matrix, and then they calculated the statistical mode of each matrix row to generate the fingerprint. Then, matching values in the fingerprints were compared according to a similarity threshold. The authors were able to identify two computer sets of 176 and 89 devices, and achieved 85\% on a web-based implementation. Compared to this work, single-board devices do not include an RTC with which to compare CPU time (two different clocks are required to analyze their deviation) and usually only contain one physical oscillator. Furthermore, after practically experimenting with this approach on single-board devices (see Section~\ref{sec:discussion}), it has been found that the resolution when measuring time on single-board devices does not allow this solution to be applied.

Additionally, some works have addressed identical device identification based on clock-skew calculated from network packets \cite{kohno2005remote,sharma2012skew,radhakrishnan2014gtid} or wireless beacons \cite{jana2009skew}. However, they have shown scalability issues when the number of devices increases and require a common observer in the fingerprint and identification process; if the observer changes, the identification is no longer possible \cite{radhakrishnan2014gtid,lanze2012skew,polcak2015clock}. Besides, raw radio frequency measurements \cite{Jafari_Fingerprinting_Deep_Learning_2018,Riyaz2018Radio} and Bluetooth transmissions \cite{huang2014blueid} have also been used to identify devices uniquely, but these methods, as other wireless-based methods, require a near physical location to the fingerprinted device. 

Based on hardware performance behavior, Wang et al. \cite{wang2012flash} analyzed the differences that occur when writing a page in a Flash chip based on manufacturing variations. To evaluate different fingerprints of the same page, the authors used Pearson correlation coefficient. Based on their experiments on 24 chips, the authors showed an estimated false positive chance of \num{4.52e-815}, and a false negative chance of \num{2.65e-539}. However, not every device includes a Flash chip to apply the technique and its usage requires knowledge of low-level hardware. Recently, Dong et al. \cite{dong2019cpg} developed a fingerprinting method based on the CPU usage graph generated while the device executes a cyclical task. The authors achieved a 93.43\% uniqueness in generated fingerprints when comparing them using the Dynamic Time Warping algorithm. However, the authors did not take into consideration critical aspects such as variable frequency or process scheduling between device cores affecting the identification stability. Regarding GPU, Nakibly et al. \cite{nakibly2015hardware} exploited GPU frequency and skew by using CPU clock as reference. Statistical fingerprints generated while rendering complex graphics show differentiation capabilities on 9 identical desktop computers, but no advanced tests were performed regarding fingerprint reliability and stability. In fact, the authors conclude that other factors to the GPU clock skew should be considered in a successful fingerprinting method.

\tablename~\ref{tab:identical_device_table} compares the main characteristics of the previous works. After reviewing these related works, the following points are extracted as conclusions. It is critical to develop modern fingerprinting mechanisms taking into account IoT device capabilities and constraints, as no previous work considered this application scenario. Besides, to ensure that the fingerprinting mechanisms are fully operative, they should be defined through a methodology able to verify that the solution is reliable and applicable in real word scenarios. 

\section{PROBLEM STATEMENT}
\label{sec:problem}

This section presents the particularities of single-board devices to later illustrate the threat model of single-board device identification. After that, it describes the properties that identification solutions based on behavioral fingerprinting should meet in the presented scenario. Finally, it highlights the limitations of existing work and motivates the necessity of novel solutions.

\subsection{Single-Board Device Description}

Although single-board devices offer great flexibility in terms of applications and operating systems, there are essential characteristics to consider before dealing with their identification. The main one is that all processing, memory, input/output, and other components are integrated into a single circuit board. In contrast, standard computers have several circuit boards for different components. This fact brings the following special aspects to consider:

\begin{itemize}
    \item \textbf{Reduced number of crystal oscillators.} Due to the objective of reducing costs, single-board computers usually dispense with components that are not critical. Thus, most devices eliminate the RTC and other physical oscillators, simulating their presence through software or using another oscillator as source frequency. The most common is to have only one or two oscillators, one for the base frequency of the processing components and another for USB and network interfaces. 
    \item \textbf{Many processing components integrated into a System on a Chip (SoC).} SoCs integrate microcontrollers with more advanced processing units such as CPUs, GPUs, or memory circuits in a single chip. As each of these components uses a different frequency to operate, it is common to use Phase-Locked Loops (PLLs) in the SoC \cite{pawar2017wide}, circuits that multiply a base frequency depending on the voltage they receive as input.
    \item \textbf{Constrained processing power}. Although single-board computers offer increasingly higher computing capabilities, they also aim to maintain low resource consumption and low price. For these reasons, the performance of single-board computers is not comparable to that of today's computers or servers. This is important and should be taken into account when generating the fingerprint.
\end{itemize}

\subsection{Threat model}
\label{sec:threat}

The main threat against the single-board device identification scenario is an adversarial actor trying to introduce a illegitimate device in a critical environment, such as an industry, by impersonating or spoofing a legitimate one. This attack could be tackled from several perspectives:

\begin{itemize}

    \item \textit{TH1. Device spoofing} \cite{marabissi2022iot}. The main security threat to solve is an adversarial entity replacing a legitimate device with a software identical malicious device. Here, the adversary uses the same legitimate software identifiers, but including malicious processes and functionality.
    
    \item \textit{TH2. Sybil} \cite{rajan2017sybil}. A single device (or many) may try to generate multiple identities to send fake data from many simulated devices. The threat of a system to Sybil attacks depends on (i) how easy the generation of identities is; (ii) whether the system treats all entities identically, and (iii) the degree to which the system accepts entries from entities that do not have a trust chain that links them to a trusted entity.  
    
    \item \textit{TH3. Advanced persistent threat} \cite{chen2022machine}. This threat arises as a consequence of the previous one. A malicious device deployed in the environment might be able to collect data from the scenario itself and from other devices, or perform further attacks such as vulnerability scan and or Denial of Service (DoS) attacks. Besides, modern attacks usually include evasion techniques that hide their activities to software-based behavior monitoring security solutions \cite{li2020adversarial}.

\end{itemize}


    


In order to solve the previous threats in this work, it is assumed that even if the device is malicious, the control over it is maintained by its legitimate administrator and the identification tasks can be executed, so that if this control is lost, it would be directly assumed that the device is infected or there is some error.

\subsection{Device identification properties}
\label{sub:properties}

In order to solve the previous threats, it is needed a proper identification mechanism able to meet properties that guarantee a consistent and reliable verification process, without forgetting the threat model depicted in Section \ref{sec:threat}. Similar properties have been defined before \cite{ruhrmair2012security}, but some of them are not suitable for IoT and single-board devices. These characteristics encompass from the fingerprint generation method to the morphology of the data generated and its manipulation. Thus, they are essential metrics to evaluate the performance of a device fingerprinting solution. In case one of them is no longer met, the solution will be severely affected in real-world deployments, limiting its usability when it comes to uniquely identify each device in the scenario.


\textbf{Uniqueness} \cite{sembiring2021randomness}. An efficient fingerprinting method should be able to uniquely identify its associated device. In other words, a fingerprint should not be generated by two different devices.

\textbf{Stability} \cite{hamza2018clear}. The fingerprint generated by a device should be consistent in time. It means that a new fingerprint of a given device should be similar enough to the previous ones of the same device.

\textbf{Diversity} \cite{ahmed2022analyzing}. The data sources and data format used to generate the fingerprint should be varied enough, so different devices generate different fingerprints. This characteristic is intrinsically related to stability, as increasing too much fingerprint diversity can affect its stability, and vice versa.

\textbf{Scalability} \cite{arellanes2020evaluating}. The fingerprint should continue being unique as the number of devices to be identified increases. This can be achieved by adding additional features to the fingerprint or by looking for features that ensure uniqueness. Thus, this characteristic is very closely related to the uniqueness property discussed before.

\textbf{Efficiency} \cite{peng2018toward}. To have a fingerprint useful for a live identification process, the generation and evaluation should not consume excessive resources, either in processing power or time.

\textbf{Robustness} \cite{zhou2019design}. The generation of the fingerprint must be immune to changes in the context that may affect the data used in the fingerprinting process. These changes in the context may include elements such as temperature, time synchronization, or resource exhaustion, among others.

\textbf{Security} \cite{lu2018internet}. The fingerprint should be secure to tackle device unauthorized access or adversarial attacks. This property implies a complete fingerprint life cycle, from its generation to storage and comparison in future identification processes.

\subsection{Limitations of existing work}

Although a good number of solutions are present in the literature, as reviewed in Section \ref{sec:related}, they are not suitable for single-board devices and the device identification task that this work pretends to fulfill. The conditions identified, which have not been covered altogether in a single work, can be summarized as:

\begin{itemize}
    \item No additional hardware or device component modification is required. In this sense, no previous solution intends to design a solution for commercial-off-the-shelf (COTS) IoT devices, where devices already available in the market do not need any modification to be physically fingerprinted.

    \item Suitable for IoT devices, specially single-board devices. Many solutions for performance-based identification leverage components such as RTCs, which are not usual in IoT devices due to their reduced price. Besides, some authors proposed intrinsic PUFs that do not require additional hardware components. However, most IoT devices include DRAM chips due to their cheaper cost, which can not be used to generate intrinsic PUFs, making these solutions unsuitable for real-world scenarios.

    \item Tested stability and robustness. Some solutions in the literature show favorable identification results. However, they do not analyze critical factors affecting performance-based identification, such as the impact of device rebooting, temperature, etc.
    
    \item Remote identification. The identifying entity does not need to be physically close to the identified device or in the same local network, as in the case of clock skew-based identification.
    
\end{itemize}

\tablename~\ref{tab:problem} evaluates the conditions correctly accomplished in each one of the works reviewed in the literature regarding individual device identification (Section \ref{sec:related}. As it can be seen, no work meets the three characteristics wanted in the present work.

\begin{table}[htpb]
    \centering
    \scriptsize
    \caption{Limitations found in each literature work.}
    \label{tab:problem}
    \begin{tabular}{ >{\Centering}m{0.9cm} >{\Centering}m{1.2cm} >{\Centering}m{1.4cm} >{\Centering}m{0.9cm} >{\Centering}m{1.0cm}  >{\Centering}m{0.9cm} }
    \hline
    \textbf{Work} & \makecell[c]{\textbf{Type}} & \makecell[c]{\textbf{No hardware}\\ \textbf{modification}} & \makecell[c]{\textbf{IoT}\\\textbf{suitable}} & \makecell[c]{\textbf{Tested}\\ \textbf{stability}} & \textbf{Remote} \\
    \hline
    \cite{salo2007multi} & Oscillator-based & \yestick & \notick & \yestick & \yestick \\
    \hline
    \cite{jana2009skew} & Clock skew & \yestick & \yestick & \notick & \notick \\
    \hline
    \cite{sharma2012skew} & Clock skew & \yestick & \yestick & \notick & \notick \\
    \hline
    \cite{wang2012flash} & Flash chip-based & \yestick  & \notick & \notick & \yestick \\
    \hline
    \cite{radhakrishnan2014gtid} & Clock skew & \yestick & \yestick & \notick & \notick  \\
    \hline
    \cite{nakibly2015hardware} & Oscillator-based & \yestick & \yestick & \notick & \yestick \\
    \hline
    \cite{Sanchez2018clock} & Oscillator-based & \yestick & \notick & \yestick & \yestick \\
    \hline
    \cite{Jafari_Fingerprinting_Deep_Learning_2018} & Radio-based & \yestick & \yestick & \notick & \notick \\
    \hline
    \cite{Riyaz2018Radio} & Radio-based & \yestick & \yestick & \notick & \notick \\
    \hline
    \cite{dong2019cpg} & CPU usage-based &  \yestick & \yestick & \notick & \yestick \\
    \hline
    \cite{babaei2019physical} & PUF & \notick & \yestick & \yestick & \yestick\\
    \hline
    \cite{kong2013processor} & CPU PUF & \notick & \yestick & \yestick & \yestick\\
    \hline
    \cite{gao2019building} & Intrinsic PUF & \yestick & \notick & \yestick & \yestick\\
    \hline
    This work & Oscillator-based & \yestick & \yestick & \yestick & \yestick \\
    \hline
    \end{tabular}
\end{table}


\begin{figure*}[ht!]
    \centering
    \includegraphics[width=\textwidth]{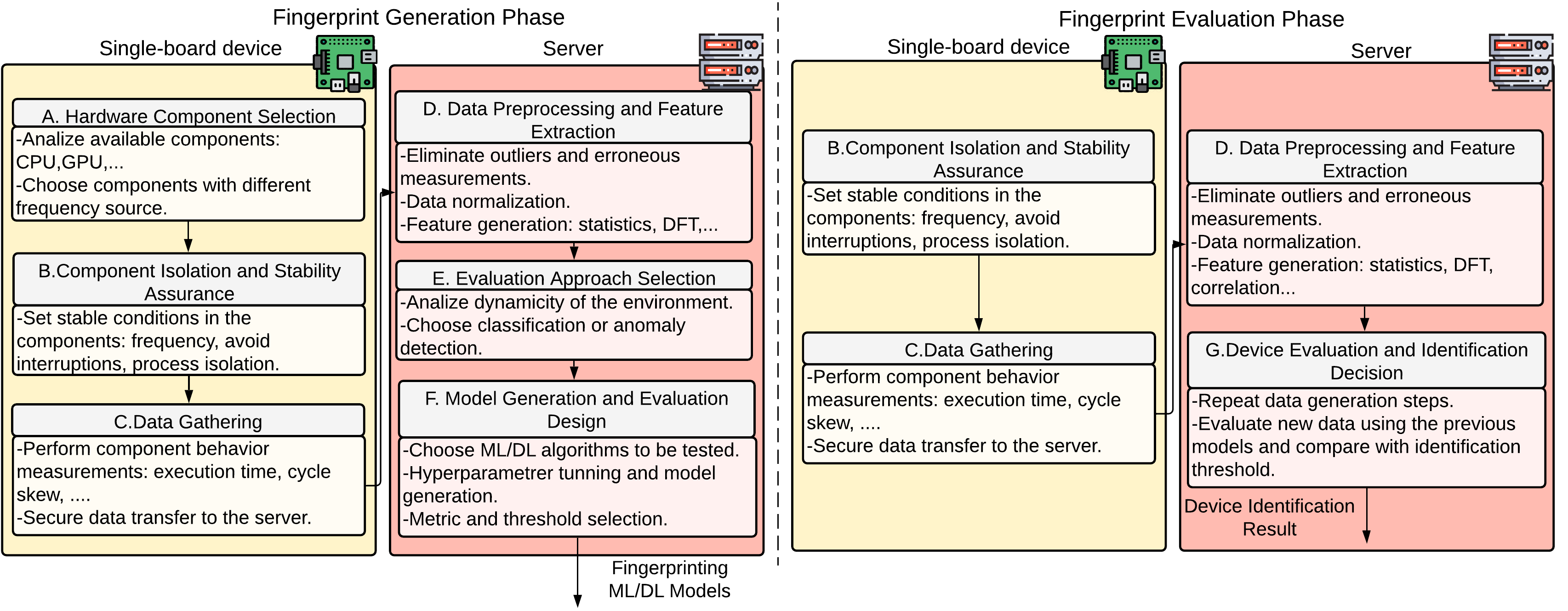}
    \caption{Graphical representation of the proposed methodology for device fingerprinting and identification.}
    \label{fig:methodology}
\end{figure*}

\section{METHODOLOGY DEFINITION}
\label{sec:methodology}

This section presents a novel methodology to identify identical single-board devices using behavioral fingerprints. It focuses on measuring the impact that insignificant hardware differences, which happened during the device manufacturing process, have on the device performance to create unique and stable behavior fingerprints. These differences are recognized by analyzing the performance of several heterogeneous components, according to parameters such as execution time or number of cycles. Thus, it is worth noting that this methodology could be applied to other types of devices containing at least two components to compare their behavior. Besides, ML/DL techniques are applied as processing tool following the best practices in the area of ML application for cybersecurity \cite{arp2022and}, but other statistical methods could also be suitable.

As shown in \figurename~\ref{fig:methodology}, the proposed methodology follows a client/server model and is composed of two fundamental phases: a first one of generation and a second of evaluation. During the fingerprint generation phase, the objective is the creation of a fingerprint per device by training ML/DL models for later device identification. During the fingerprint evaluation phase, new fingerprints per device are generated to be evaluated with the ML/DL models trained in the previous phase, giving a final identification output for the device. The methodology consists of the next seven fundamental steps, which can be repeated in both phases depending on the tasks to be carried out:

\begin{itemize}
    \item (A) \textit{Hardware Component Selection}. Select the device components whose behavior is going to be analyzed.
    \item (B) \textit{Component Isolation and Stability Assurance}. Establish stable conditions for the components, reducing external inferences to a minimum.
    \item (C) \textit{Data Gathering}. Measure the behavior of device hardware components.
    \item (D) \textit{Data Preprocessing and Feature Extraction}. Remove erroneous measurements, normalizes them, and extracts new significant values.
    \item (E) \textit{Evaluation Approach Selection}. Decide between classification or anomaly detection depending on the environment properties.
    \item (F) \textit{Model Generation and Evaluation Design}. Train ML/DL algorithms, select performance metrics, and establish model thresholds.
    \item (G) \textit{Device Evaluation and Identification Decision}. Repeat steps B, C, and D to perform device identification.

\end{itemize}

\begin{figure}[ht!]
    \centering
    \includegraphics[width=0.9\columnwidth]{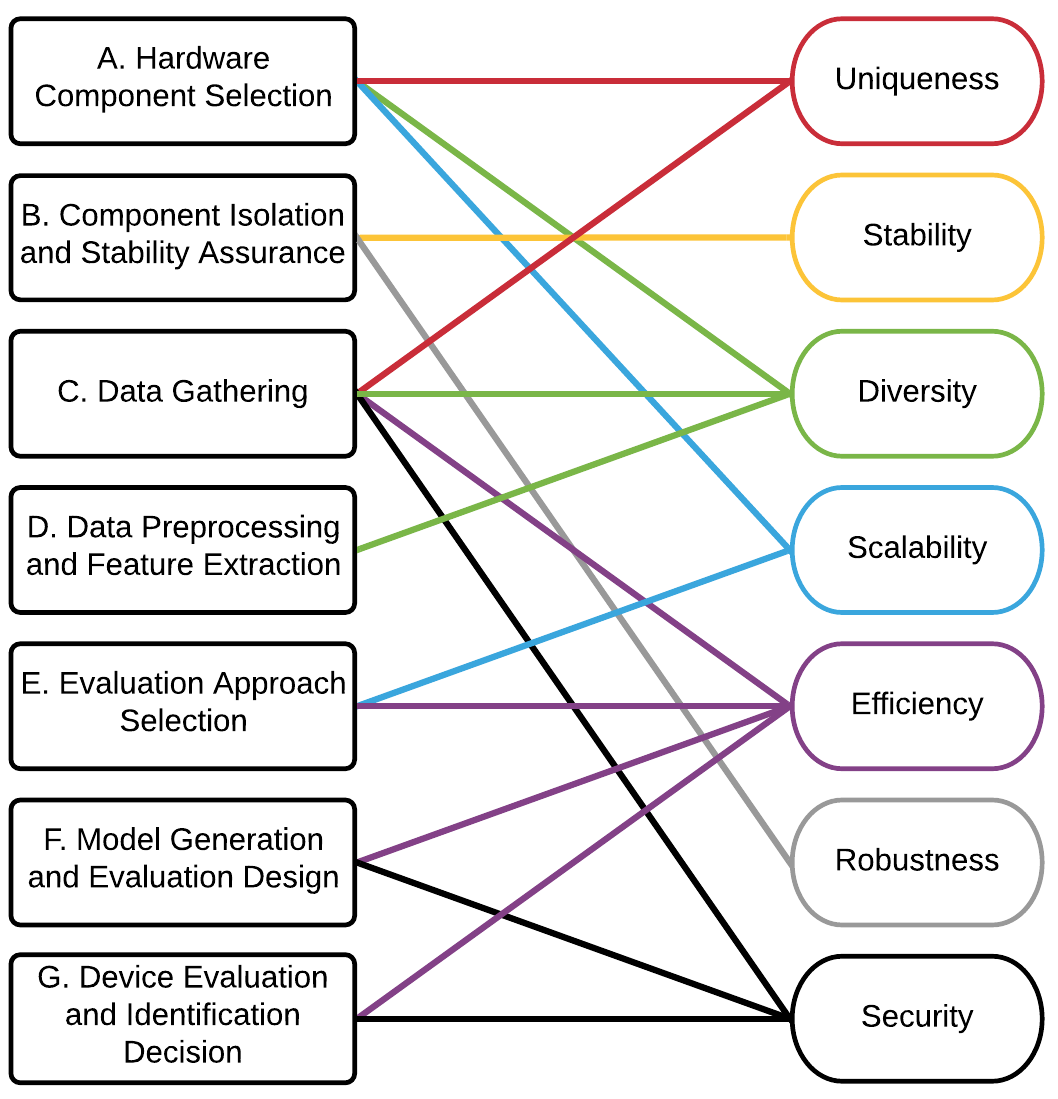}
    \caption{Association between methodology steps and fingerprint properties.}
    \label{fig:properties_steps}
\end{figure}

\figurename~\ref{fig:properties_steps} shows the relationship between the different steps detailed above and the properties desired in an individual device identification solution, as introduced in Section \ref{sub:properties}.

\subsection{Hardware Component Selection}

The first step is to analyze the hardware of the device where the fingerprint needs to be generated. The goal is to identify components with potential manufacturing variations whose performance can be accurately measured and compared. 

In this sense, since the fingerprint will be based on device self-contained hardware, it is necessary to identify at least two components to be used, as their behavioral performance will be compared to each other since one component can not notice its own performance imperfections without a reference point, although to improve the \textit{scalability} and \textit{diversity} of the fingerprint more could be added if available. The preference here is to select components whose frequency is based on different physical oscillators, as their differences will be larger, although components with different frequencies sharing one oscillator as the base frequency can also be compared. Examples of components to consider in single-board devices: CPU, GPU, memory, network controllers, USB controllers, or time control oscillators.

\subsection{Component Isolation and Stability Assurance}

Once the hardware components to be monitored are chosen, the next step is to establish a configuration that ensures the \textit{stability} of the behavioral measurements. This step seeks to ensure a stable and identical condition during the generation of the fingerprint, both for training and testing phases. At this point, it is critical to guarantee that there are no external elements introducing noise or variability.

With that goal in mind, one of the key factors to take into account is the frequency at which the component is operating, since in single-board computers it is common for the operating system to establish some adaptability according to the load on the system or the need to save energy. Thus, it is necessary to ensure that the fingerprint will be generated under identical frequency conditions. Otherwise, it would be impossible to compare the variation in performance between various components. In this sense, components such as the CPU or GPU are the ones that can have more variability in their operating frequency, ranging from some MHz when are in power-save mode to several GHz when they are under high-performance requirements. Another aspect to take into account is the isolation of the software that performs the measurements with respect to other programs running on the system. The measures to guarantee this isolation include the separation of some of the CPU cores from the general process scheduler, the use of transactional memory \cite{harris2010transactional}, the disabling of interrupts by the kernel or isolating the GPU. Note that the exact actions may vary according to the components chosen. Moreover, it is also important to control external conditions such as temperature to the extent possible, since it can influence the performance variation of some components. In the case of using CPU timers, time synchronization made by services such as NTP should be also considered. These considerations seek to improve the \textit{robustness} of the fingerprinting solution.

\subsection{Data Gathering}

When the desired stability conditions have been achieved, it is necessary to define the functions to be performed on the components (selected in phase A) to measure their behavior in parallel and determine the possible skew between them. In this sense, the measurements must allow the comparison of the performance of two different components from the same device, avoiding executing the operations and measuring the deviation using a unique component.

Choosing the functions to run on each component to compare their behavior is a critical task during the fingerprinting solution design and must be carefully studied to ensure the \textit{efficiency}, \textit{diversity}, and \textit{uniqueness} of the fingerprint. Due to the fact that functions taking longer times to execute may better show the variance between components, but may make the fingerprint generation process take too long. In addition, the created approach should not consume too many resources as it could slow down the normal operation of the system and affect other tasks. For example, the authors of \cite{Sanchez2018clock} decided to measure functions that take a short time to execute using the RTC, comparing CPU, and RTC oscillators. Besides, the authors of \cite{salo2007multi} measured the clock cycles in one second compared with the RTC and when processing one second of audio using the DSP. Moreover, to fulfill the \textit{security} property, the data collection process should be executed using Trusted Execution Environments (TEEs) \cite{lee2020softee}, if available, to isolate the fingerprint generation task from the rest of the device processes. This avoids possible data leaks caused by attacks based on memory vulnerabilities. Finally, the generated behavioral data is sent to a server, where it will be processed to generate the fingerprint. This sending should be done over a secure communications channel, such as SSH or TLS, avoiding possible interceptions of data transmissions.

\subsection{Data Preprocessing  and Feature Extraction}

Once the server receives the behavioral data, the next step is to preprocess the data to eliminate possible erroneous measurements and extract new information. Here, note that the data gathering step could be done several times before going to data preprocessing and feature extraction. The final objective is the generation of a set of feature vectors that will act as the generated fingerprint data, guaranteeing the \textit{diversity} in the values. These vectors are the ones that will later be used to feed the ML/DL algorithms and generate the models.

To start the preprocessing part, it is necessary to remove outliers, constant, corrupted, or missing values that may be in the dataset by scanning over the dataset. For that, it is useful to plot each set of values collected and remove values that are more than 3 standard deviations away from the average. Afterwards, it is highly recommended to scale the data and have it in the same data range. In this sense, the most common scaling algorithms are min-max and standard normalization.

Once errors have been removed and the values scaled, the next step is feature extraction. It is possible to extract different features from the series of values by grouping them together and calculating different characteristics of the resultant series. One of the most typical values is the extraction of statistical values such as mean, median, deviation, max, min, or mode. However, applying more advanced calculations can provide even more relevant information about possible latent features in the values. In this sense, Discrete Fourier Transform (DFT) or Discrete Wavelet Transform (DWT) can be applied to extract features related to the time and order of the values. In addition, it is also possible to calculate features based on the correlation between the available values using algorithms like the Pearson correlation. Associating this step with works in the literature, the authors of \cite{Sanchez2018clock} used the mode of a series of 1000 values, and the authors of \cite{BabunCPS2021} and \cite{wang2012flash} employed the average of the measurements taken and the correlation in the generated values.
    
\subsection{Evaluation Approach Selection}

Once the features that will generate the fingerprint have been obtained, it is necessary to define the ML/DL approach to be followed \cite{cadavid2020machine}. There are two possibilities here, a classification approach, in which the different devices in the environment will be associated with a label, and an anomaly detection approach, where the data from each device is labeled as ``normal'' and a separate model is generated for each of them. 

This decision must be made taking into account both the scenario (number of devices, variety of devices, possibility of adding or removing devices) and the features that have been collected (similarity of values between devices, number of features, etc.). Thus, an environment with a low number of devices may benefit from the use of classification algorithms, while more dynamic environments with a large number of devices will need more varied features and will benefit from anomaly detection algorithms. Here, the \textit{scalability} and \textit{efficiency} of the approach are better if no retraining is needed each time a device joins or leaves the scenario. In the literature, solutions have been found with both approaches, applying classification perspectives \cite{BabunCPS2021} or generating a statistical model per device and confronting the new fingerprints to it when identification is to be performed \cite{Sanchez2018clock}.

\subsection{Model Generation and Evaluation Design}

Once the desired approach has been selected, either classification or anomaly detection, it is necessary to train ML/DL algorithms and define the metrics that will be used in the identification. This step should be carried out considering the \textit{efficiency} in the evaluation process and the \textit{security} against possible data-based attacks to the models.

There is a wide variety of algorithms that can be considered in this step, differentiating between traditional ML algorithms and DL algorithms based on neural networks \cite{cadavid2020machine}. Starting from classification, algorithms such as Random Forest, k-Nearest Neighbors, eXtreme-Gradient Bosting (XGBoost), Support Vector Machines (SVM), or Multi-Layer Perceptron (MLP) can be used. From the anomaly detection prism, Isolation Forest, Local Outlier Factor (LOF), One Class-SVM, or Autoencoders are good alternatives as well. At this stage, it is also worth considering the application of algorithms focused on time series \cite{cadavid2020machine}, depending on whether there are time-based dependencies between the values. Once the algorithms to use have been selected, it will be necessary to train and fine-tune the hyperparameters that give the best results in each of them. Note that these hyperparameters will vary according to the selected algorithms. In addition, the model predictions are usually one per vector, so they cannot be used directly to give a decision during the evaluation and identification of the device. In this sense, it is common to determine a \textit{threshold} based on the model performance from which the device under evaluation will be accepted as the legitimate one. This threshold can be defined using numerous equations or conditions, such as defining the 50\% of the values being recognized as legitimate, as done by the authors in \cite{Sanchez2018clock}. Common metrics to consider on this step are \textit{accuracy}, \textit{true positive rate (TPR)}, \textit{false positive rate (FPR)}, or \textit{F1-Score}, among others \cite{cadavid2020machine}.

At this point, it is worth noting that although this methodology has been designed primarily for ML/DL algorithms due to their current prominence in many research fields, it could be possible to include in this step other statistical algorithms, or even some self-developed algorithms as in \cite{BabunCPS2021}.

\subsection{Device Evaluation and Identification Decision}

This step is only carried out in the evaluation phase and involves generating new behavioral data of the device following the same methodology as during the training phase, repeating steps B, C, and D.

Once the new dataset is generated, it is used to identify the device, determining whether it is the same device used during training or not. To this end, data will be evaluated using the ML/DL models previously generated, so that one result per vector is obtained. Then, the rule determined in the previous step will be applied, either based on a threshold or another equation to give a final decision on the device identification.

\section{METHODOLOGY VALIDATION}
\label{sec:PoC}


This section validates the suitability of the proposed methodology by implementing a Proof-of-Concept (PoC) on a realistic scenario composed of 20 identical single-board devices. In particular, the devices are 15 Raspberry Pi 4 Model B 2GB (RPi 4) and 10 Raspberry Pi 3 Model B+ (RPi3) running identical software images, with Raspbian 10 (buster) as OS and 5.4.83 as Linux kernel version. The operating systems ran in head-less mode, i.e., without a graphical environment or output to a display, a common configuration in SOC devices deployed in IoT. Next, it is detailed how the methodology has been implemented in the previous scenario, describing the decisions made in each of the defined steps. The language used has been Python and the code is available in \cite{code}, for reproducibility sake.

\begin{figure*}[htpb]
	\centering
	\begin{subfigure}{0.45\textwidth}
    \includegraphics[width=\textwidth,trim={25 0 30 0} ,clip=true]{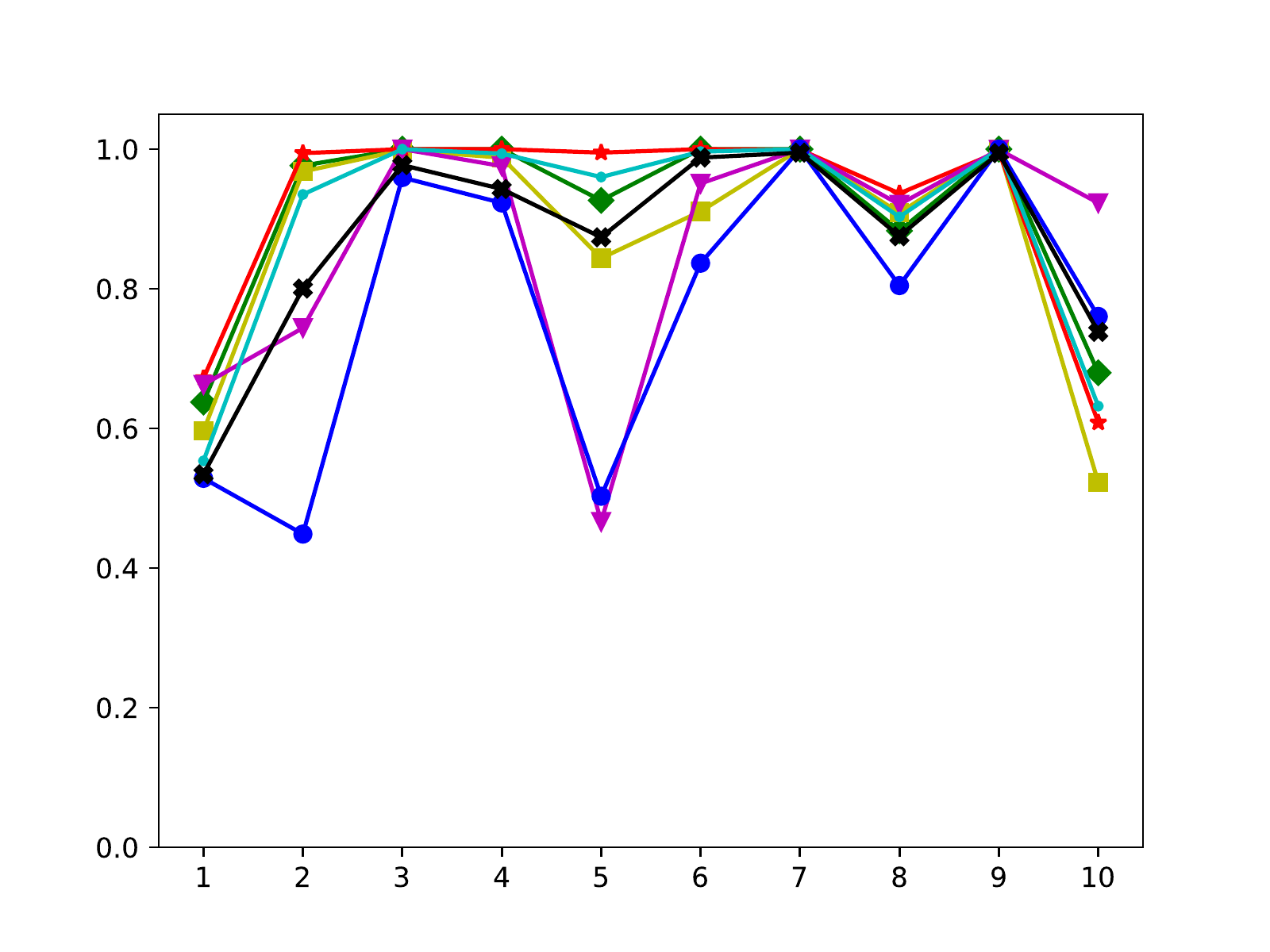}
    \caption{Average TPR for each RPi3.}
    \end{subfigure}
	\begin{subfigure}{0.47\textwidth}
	\includegraphics[width=\textwidth,trim={25 0 30 0} ,clip=true]{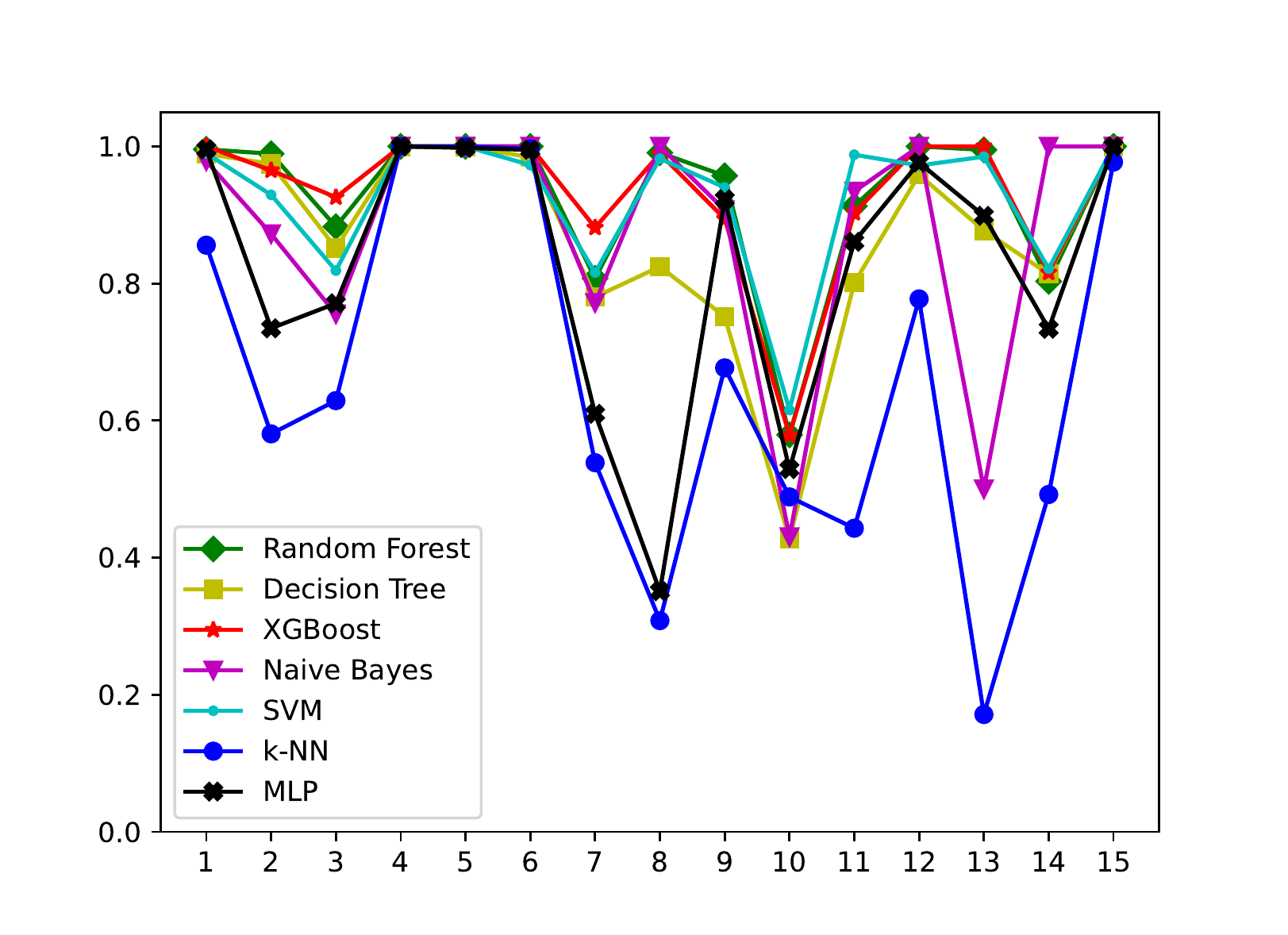}
	\caption{Average TPR for each RPi4.}
	\end{subfigure}
    \caption{Fingerprint evaluation TPR during validation per device and model.}
    \label{fig:res}
\end{figure*}

\textit{A. Hardware Component Selection.} 
As a starting point, the physical oscillators available in the RPi4 and RPi3 were analyzed. The result of this study concluded that one oscillator is shared between the SoC components, running at 54 MHz in RPi4 and 19.2 MHz in RPi3, and the USB controller running at 25 MHz in both models \cite{undocumentedPi}. Since accessing the frequency of the USB oscillator from the device is not simple, the selected components were the VideoCore VI GPU and the ARM Quad-core Cortex-A72 CPU for RPi4 and VideoCoreIV GPU and the ARM Quad-core Cortex-A53 for RPi3. Although they share the base oscillator (GPU and CPU), their frequencies are given by different PLLs.

\textit{B. Component Isolation and Stability Assurance}. 
Both the CPU and GPU work at varying frequencies depending on the load on the device. So, to guarantee the stability of the signatures, it is needed to ensure that frequency is fixed. For the validation, the RPi4 CPU frequency was set to 1.5 GHz and the GPU one to 500 MHz, while the RPi3 CPU frequency was set to 1.4 GHz and the GPU one to 400 MHz, the maximum values of both by default (without overclock). To do this, the Turbo Mode was enabled by adding \textit{force\_turbo=1} in \textit{/boot/config.txt}. After that, one of the CPU cores was isolated to be used in the fingerprint generation, using the options in \textit{/boot/cmdline.txt}, preventing processes from being assigned to it.

\textit{C. Data Gathering}. 
To measure the variation of behavior between components, it was compared how the cycle counters of each component (CPU and GPU) vary with respect to the other. To do this, \textit{sleep}, \textit{random number generation}, and \textit{hash} functions were selected. In particular, these functions were sequentially executed in the CPU and the number of GPU cycles that occurred during each function execution was measured. To interact with the GPU, Idein's py-videocore6 library \cite{py-videocore6} was used in RPi4. Concretely, the CORE\_PCTR\_CYCLE\_COUNT GPU counter was the register monitored. In the case of RPi3, Idein's py-videocore \cite{py-videocore} library was used to monitor the QPU\_Total\_idle\_clock counter (as the RPi3s were in headless mode). The data gathering procedure is summarized in \algorithmcfname~\ref{alg:gathering}. For the data collection, the sleep function time \textit{t} was set to 120 seconds, as the variations between CPU and GPU are presumably low, a fixed string was set for the hash function, and the number on measurements (\textit{n\_measurements}) was set to 400. It is important mentioning that these values were adjusted according to the results in later steps. Other configuration parameters such as \textit{t=60} seconds were tested providing with slightly worse results. Additionally, the use of TEE to run the algorithm was considered, however the ARM TrustZone instance available in RPi is simulated only \cite{op-tee}.

\begin{algorithm}[ht!]
\SetAlgoLined
\KwResult{Set of GPU/CPU performance measurements. }
 result\_set=\{\}\;
 \For{n in n\_measurements}{
  \#Sleep cycle counter\\
  GPU\_CYCLE\_COUNT=0;\\
  sleep(t);\\
  sleep\_gpu\_cycles=GPU\_CYCLE\_COUNT;\\
  \vspace{0.2cm}
  \#Random number generator cycle counter\\
  GPU\_CYCLE\_COUNT=0;\\
  random\_number\_gen();\\
  random\_gpu\_cycles=GPU\_CYCLE\_COUNT;\\
  \vspace{0.2cm}
  \#Hash cycle counter\\
  GPU\_CYCLE\_COUNT=0;\\
  hash("Test string");\\
  hash\_gpu\_cycles=GPU\_CYCLE\_COUNT;\\
  \vspace{0.2cm}
  \#Add measurements to result set\\
  result\_set.append("sleep\_gpu\_cycles, random\_gpu\_cycles,hash\_gpu\_cycles");\\
 }
 \caption{CPU/GPU data acquisition algorithm}
 \label{alg:gathering}
\end{algorithm}

\textit{D. Data Preprocessing and Feature Extraction}. 
The data gathering process was repeated a total of ten times per device, for testing purposes, with different temperature conditions and performing several reboots between the generation of each fingerprint (set of measurements). Then, the 400 measurements of each fingerprint were grouped in different sliding windows ranging from 10 to 100 values in jumps of 10 values (10 different sliding windows in total). Afterwards, several statistical features were calculated for each window-based group and concatenated together. Concretely, the statistical values calculated were: \textit{minimum}, \textit{maximum}, \textit{mean}, \textit{median} and \textit{sum}. Following this approach, the resultant vectors for training and evaluation have a size of 150 (3 data gathering functions * 10 different sliding windows * 5 statistical features).



\textit{E. Evaluation Approach Selection}. 
Due to the staticity of the test environment, as the number of devices do not change in time, it was decided to follow an approach based on classification ML algorithms combined with a threshold on the True Positive Rate (TPR) that would delimit the minimum number of successfully classified vectors. Besides, F1-Score is also calculated to validate the classification performance of the models. (TP: True Positive, FP: False Positive, TN: True Negative, FN: False Negative).

\begin{equation}
    TPR / recall=\frac{TP}{TP+FP}
\end{equation}
    
\begin{equation}
    F1-Score=\frac{TP}{TP+\frac{1}{2}(FP+FN)}
\end{equation}

\textit{F. Model Generation and Evaluation Design}. 
For the model generation, six fingerprints of each device were used as separate training in order to have cross-validation. The selected algorithms were Random Forest, Decision Tree, k-NN, XGBoost, Naive Bayes, SVM and MLP. After hyperparameter optimization (see \tablename~\ref{tab:clasif_alg_hyp}), using cross-validation with the fingerprints used for training, the best performing algorithm was XGBoost (\textit{lr=0.1, max\_depth=20, gamma=0.01, colsample\_bytree =0.5}), giving an average TPR of 91.92\%, ranging from 100\% in the best case to 55\% in the worst (a random predictor would give 4\% for each device, as the model can be easily identified based on device frequency). \figurename~\ref{fig:res} shows the results per algorithm and device. This value varies highly, as some of them seem to be more similar between them. Based on the previous results, a threshold of 50\% in the assigned classes in evaluation can be defined to give the identification decision, so that if half of the vectors are correctly evaluated, the device is recognized as legitimate.

\begin{table}[t]
	\caption{Classification algorithms and hyperparameters tested.}
	\centering
    \scriptsize
    \begin{tabular}{m{1cm}m{5.8cm}m{0.8cm}}
        \hline
        \textbf{Model} & \textbf{Hyperparameters tested} & \textbf{\makecell[l]{Avg\\TPR}}\\
        \hline
        Naive Bayes & No hyperparameter tunning required & 87.29\% \\
        \hline
         k-NN & $k\in [3,20]$ & 71.40\% \\
         \hline
         SVM & \makecell[l]{$C\in [0.01,100], gamma\in [0.001,10]$\\$kernel\in \{'rbf', 'linear', 'sigmoid','poly'\}$} & 89.65\%\\
         \hline
         XGBoost & \makecell[l]{$lr \in [0.01,0.3], max\_depth \in [3,15]$\\ $min\_child\_weight\in [1,7], gamma \in [0,0.5],$\\$colsample\_bytree \in [0.3,0.7]$} & 91.92\% \\
         \hline
         Decision Tree & \makecell[l]{$max\_depth \in [None, 5, 10, 15, 20]$ \\ $min\_samples\_split \in [2,3,4,5]$} & 86.47\% \\
         \hline
         \makecell[l]{Random\\Forest} & \makecell[l]{$number\_of\_trees \in [50,1000]$\\$max\_depth \in [None, 5, 10, 15, 20]$\\$min\_samples\_split \in [2,3,4,5]$} & 91.64\% \\
         \hline
         MLP & \makecell[l]{$layers\in [1,3], neurons\_layer\in [10,100], $ \\ $batch\_size \in [32,128,256,512]$} & 85.32\% \\
         \hline
    \end{tabular}
    \label{tab:clasif_alg_hyp}
\end{table}{}

\begin{figure}[htpb!]
    \centering
    \includegraphics[width=\columnwidth,trim={8 8 4 4},clip=true]{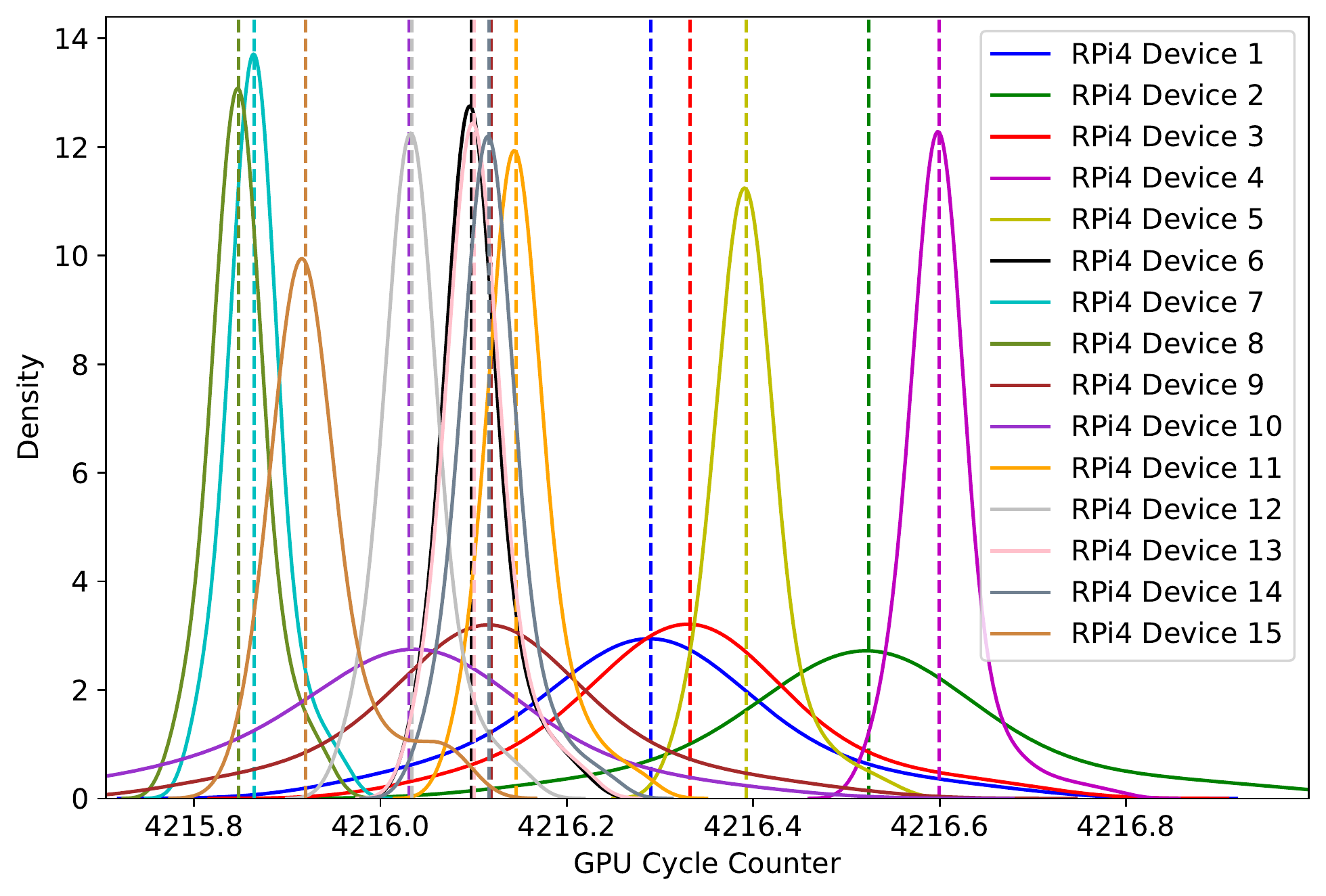}
    \caption{Density plot for the GPU cycle counter in RPi4.}
    \label{fig:density_plot}
\end{figure}

As \figurename~\ref{fig:res} depicts, the performance of the classifiers when identifying the devices is not homogeneous and they are able to classify better some devices than others. To explore more in detail this aspect, \figurename~\ref{fig:density_plot} shows the density plots for the CPU and GPU skew after the sleep function execution. In the vertical dotted line, the median of the distribution is shown. It can be appreciated how the skew of some devices varies, being more similar between some of them. Concretely, for the RPi4 number 10, the one with the worst classification performance in \figurename~\ref{fig:res}, it can be seen in \figurename~\ref{fig:density_plot} how the median of its distribution is almost identical to the RPi4 number 12. Although only one of the three functions executed is plotted due to space constraints, this analysis demonstrates how some devices are more similar to each other than others, a factor that influences the scalability of the solution, so that for larger deployments, a greater number of functions or data sources would be necessary.

\begin{figure}[t]
	\centering
	\begin{subfigure}{0.49\textwidth}
    \includegraphics[width=\textwidth,trim={95 62 220 50} ,clip=true]{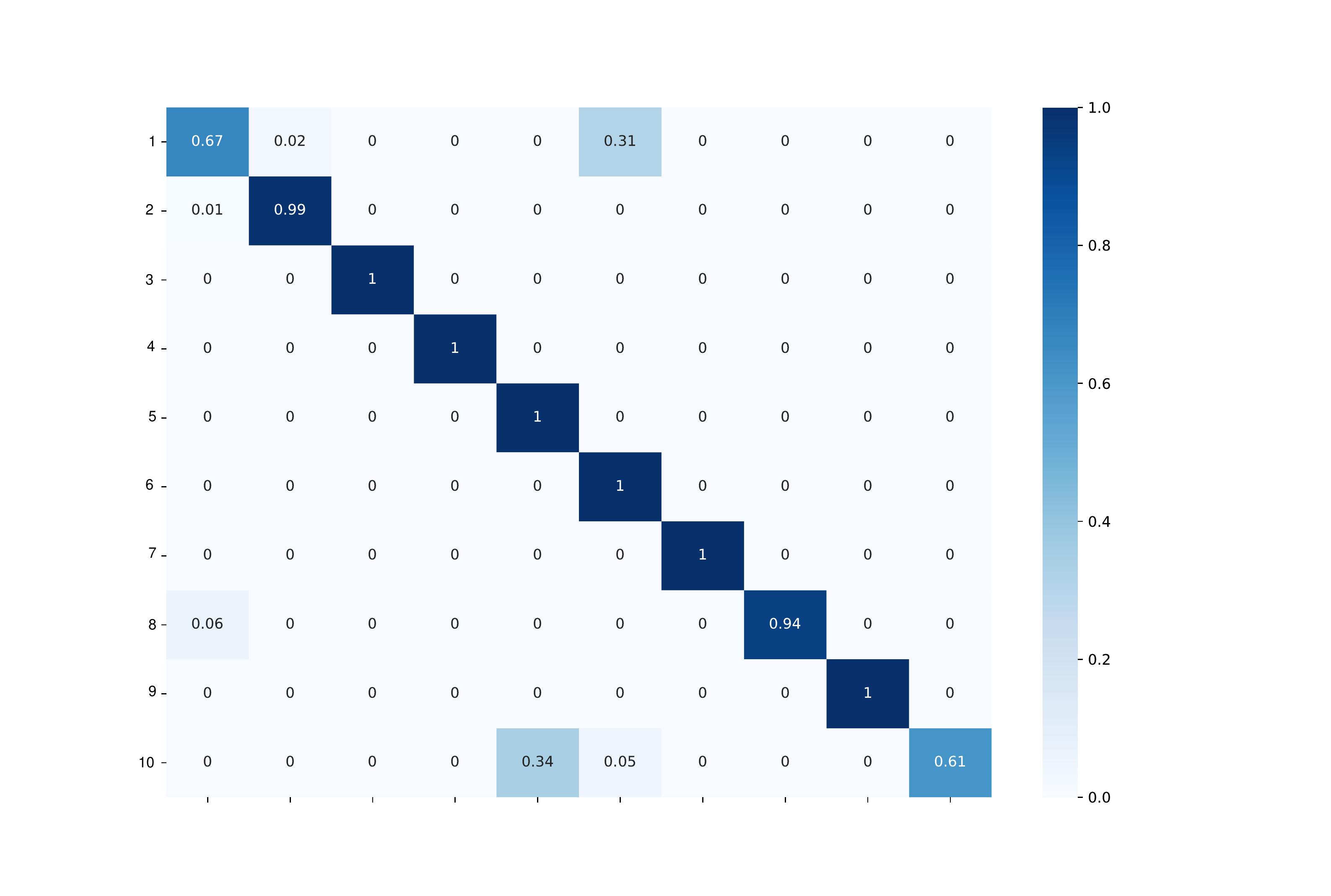}
    \caption{10 RPi3 classification confusion matrix.}
    	\end{subfigure}
	\begin{subfigure}{0.50\textwidth}
	\includegraphics[width=\textwidth,trim={110 0 299 50} ,clip=true]{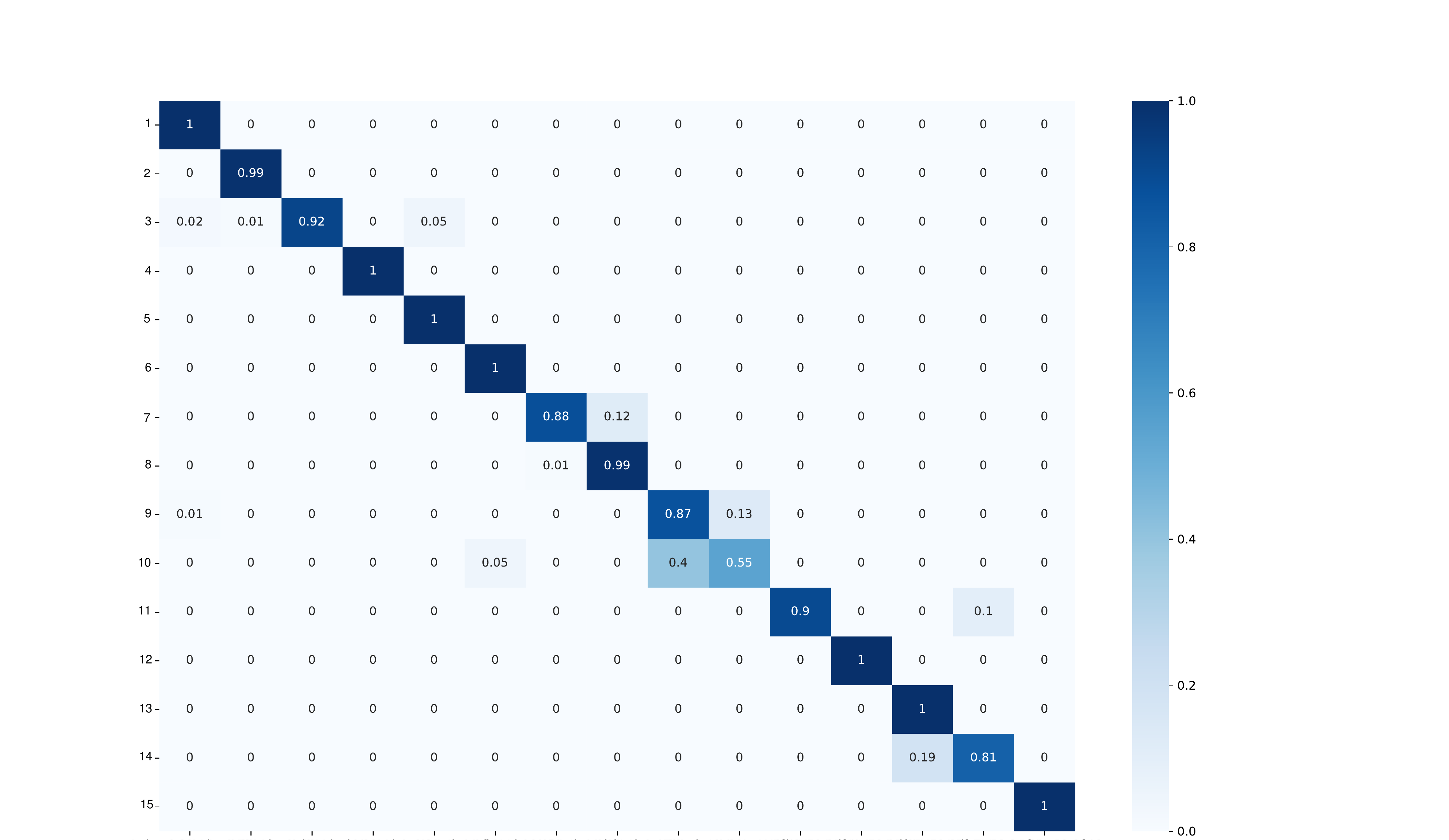}
	\caption{15 RPi4 classification confusion matrix.}
	\end{subfigure}
    \caption{Test confusion matrix for device identification using XGBoost.}
    \label{fig:matrix}
\end{figure}

\textit{G. Device Evaluation and Identification Decision}. 
In the present PoC, this phase was performed with the four fingerprints of each device not used for the previous phase. In this step, the normalization was repeated with the same values used to generate the model, and the vectors containing the same features were evaluated using the XGBoost model trained previously. Using the 50\% threshold as explained above, all the devices were correctly identified without any device erroneously identified as another one. \figurename~\ref{fig:matrix} shows the average confusion matrix for the four fingerprints used for testing, using XGBoost as classifier. The labels are defined as the device model followed by its MAC address. The evaluation is done by grouping together devices within the same model, as RPi3 and RPi4 have different running frequencies in the components leveraged and they can be easily differentiated. $\approx$93\% and $\approx$92\% average F1-score is achieved for RPi4 and RPi3, respectively.

As conclusion, it has been demonstrated the performance of the proposed methodology in an environment with real devices. Still, this is only a PoC and its performance could be substantially improved by extracting other data from devices and generating more elaborate features.

\begin{table*}[htpb]
    \centering
    \scriptsize
    \caption{Analogy between hardware-based fingerprinting solutions in the literature and the proposed methodology.}
    \begin{tabular}{ >{\Centering}m{0.6cm} >{\Centering}m{1.0cm}  >{\Centering}m{1.9cm} >{\Centering}m{2.3cm} >{\Centering}m{2.05cm} >{\Centering}m{1.35cm} >{\Centering}m{2.5cm} >{\Centering}m{3.0cm}} 
    \hline
    \textbf{Work} & \textbf{Step A} & \textbf{Step B} & \textbf{Step C} & \textbf{Step D} & \textbf{Step E} & \textbf{Step F} & \textbf{Step G} \\
    \hline
    \cite{salo2007multi} & RTC, DSP, CPU & - & CPU cycles in one second (measured with DSP and RTC) & Raw values & - & Statistical (t-test), $p<=0.05$ threshold  & 98.7\%-93.3\% identification\\
    \hline
    \cite{Sanchez2018clock} & RTC, CPU & Transactional memory & Execution time of short functions & Mode-based matrix  & - & Statistical comparison, 50\% similarity threshold & Correct identification \\
    \hline
    \cite{wang2012flash} & Flash memory & Isolation of one page in flash memory & Bit programming errors (flip from 1 to 0) & Error order per bit & - & Pearson correlation, 0.5 threshold & Estimated \num{4.52e-815} FPR and \num{2.65e-539} FNR\\
    \hline
    \cite{dong2019cpg} & CPU & Thread affinity & CPU usage while executing a cyclical tasks & Raw values & - & Dynamic Time Warping algorithm, 0.3244 threshold & 93.43\% uniqueness (Shannon entropy) \\
    \hline
    \cite{nakibly2015hardware} & CPU, GPU & - & Number of frames per 5 seconds & Entropy and statistics &  - & Statistical & No evaluation, partial differentiation capabilities \\
    \hline
    This work & CPU and GPU & Core isolation, Fixed frequency & Sleep for 120 secs, Random num. gen., hash & Sliding window-based statistical features &  Classification & XGBoost, 50\% threshold & Perfect Identification (91.92\% avg. TPR)\\
    \hline
    \end{tabular}
    \label{tab:discussion}
\end{table*}

\begin{table*}[htpb]
    \centering
    \scriptsize
    \caption{Comparison of the validation approaches implemented.}
    \begin{tabular}{ >{\Centering}m{1.6cm} >{\Centering}m{0.9cm}  >{\Centering}m{1.5cm} >{\Centering}m{2.0cm} >{\Centering}m{1.85cm} >{\Centering}m{1.35cm} >{\Centering}m{1.5cm} >{\Centering}m{2.0cm}  >{\Centering}m{1.8cm}} 
    \hline
    \textbf{Approach} & \textbf{Step A} & \textbf{Step B} & \textbf{Step C} & \textbf{Step D} & \textbf{Step E} & \textbf{Step F} & \textbf{Step G} & \textbf{Properties not met}\\
    \hline
    \cite{dong2019cpg}-inspired approach & CPU & Thread affinity & Short functions & Raw values & Anomaly Detection & LOF, 50\% threshold & Identification until device reboots (69.4\% avg. TPR) & Stability \\
    \hline
    \cite{nakibly2015hardware}-inspired approach & CPU and GPU & - & Different GPU operations & Raw values & Anomaly Detection& LOF, 60\% threshold & Identification until device reboots (89.6\% avg. TPR) & Stability\\
    \hline
    \cite{Sanchez2018clock}-inspired approach A & CPU & - & Short functions & Window-based statistical features & Classification & XGBoost, 50\% threshold & No identification (27.5\% avg. TPR)  & Uniqueness, Diversity, Stability \\
    \hline
    \cite{Sanchez2018clock}-inspired approach B & CPU & - & Short functions & Window-based statistical features & Anomaly Detection & LOF, 50\% threshold & No identification (19.8\% avg. TPR) & Uniqueness, Diversity, Stability \\
    \hline
    This work (Sec. \ref{sec:PoC}) & CPU and GPU & Core isolation, Fixed frequency & GPU-measured CPU operations & Window-based statistical features &  Classification & XGBoost, 50\% threshold & Perfect Identification (91.9\% avg. TPR) & - \\
    \hline
    \end{tabular}
    \label{tab:validation}
\end{table*}

\section{DISCUSSION}
\label{sec:discussion}

This section compares the proposed methodology with the solutions available in the literature. After that, it discusses the limitations of the proposed solution and provides some lessons learned.

\subsection{Literature comparison}

Despite the solutions discussed in Section \ref{sec:related} do not follow a common methodology, many of them implement certain steps proposed in this work. \tablename~\ref{tab:discussion} compares the proposed methodology and related work using on-device components for identification. As can be seen, all works performing identification utilize a threshold (Step F), defined based on different statistical approaches. Besides, none of the approaches employed ML/DL algorithms (Step F) and many of them did not consider hardware isolation properly or the usage of fixed component frequencies (Step B).

After the theoretical comparison, it is relevant to analyze the most similar and comparable solutions from a common prism. Although most of the solutions analyzed in Section \ref{sec:related} use components that are not available on the RPi4 or RPi3, three solutions, \cite{dong2019cpg}, \cite{nakibly2015hardware} and \cite{Sanchez2018clock}, can be adapted to the present scenario and methodology. \tablename~\ref{tab:validation} compares the methodology approach and the results of its validation with four implementations inspired by the works found in Section \ref{sec:related}. Besides, it highlights which fingerprint properties were not met, resulting in erroneous device identification.

The first of these approaches was inspired by \cite{dong2019cpg}, only the CPU was selected as a component but making the fingerprint of each of its cores separately by using thread affinity. The features to be obtained were statistics based on the time taken to perform small operations (string hash and random number generation) on each of the cores. Using LOF as an anomaly detection algorithm and one model per device, the identification was possible by setting a threshold of 50\%. However, the reboot of the devices caused the fingerprints to change and it was not possible to perform the identification due to the new kernel process scheduling, something that may also be affecting the proposed solution in \cite{dong2019cpg}. The same problem occurred in a second tested approach inspired by \cite{nakibly2015hardware}. In particular, each CPU core was compared with the GPU separately in a concurrent manner and executing short operations. Here, different operations of variable complexity were performed in the GPU while the execution time was measured using the CPU. In this case, the evaluation also followed an anomaly detection-based approach, being LOF the algorithm with the better results. Again, it was possible to identify the devices consistently, now using a threshold around 60\%, until they are rebooted. Finally, two different approaches were tested inspired by \cite{Sanchez2018clock}. Both share the fact that the data collected was based on short functions executed in the CPU without considering stability measurements. They differ in the evaluation approach, one using anomaly detection and the other using classification. These approaches achieved the worse performance, as the solutions could not identify the devices even without rebooting them.

\begin{figure*}[ht!]
    \centering
    \begin{subfigure}{0.45\textwidth}
    \includegraphics[width=\textwidth,trim={0 0 0 0} ,clip=true]{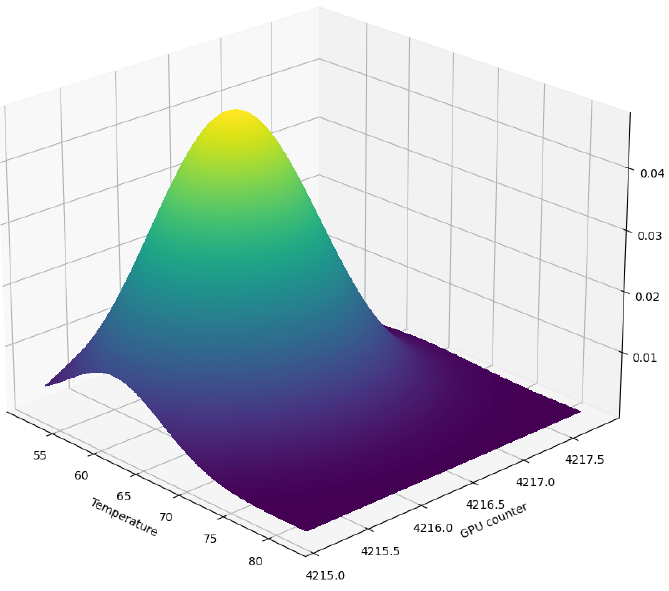}
    	\caption{Raspberry PI MAC dc:a6:32:4c:9a:79}
	\end{subfigure}
	    \begin{subfigure}{0.45\textwidth}
    \includegraphics[width=\textwidth,trim={0 0 0 0} ,clip=true]{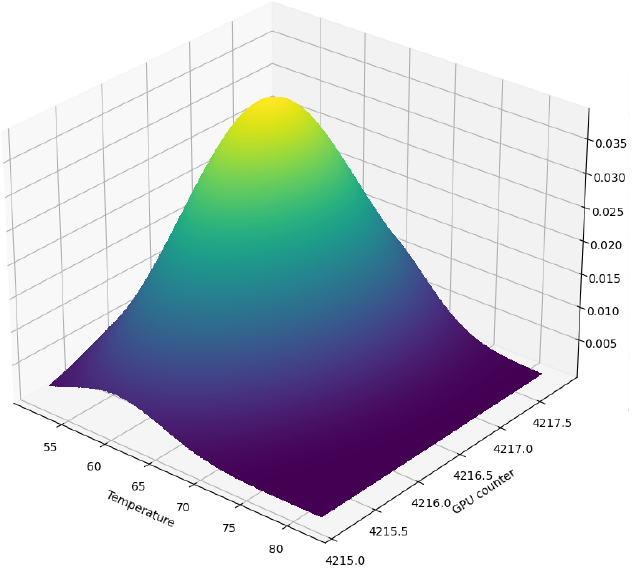}
    	\caption{Raspberry PI MAC dc:a6:32:4c:90:fb}
	\end{subfigure}
	    \begin{subfigure}{0.45\textwidth}
    \includegraphics[width=\textwidth,trim={0 0 0 0} ,clip=true]{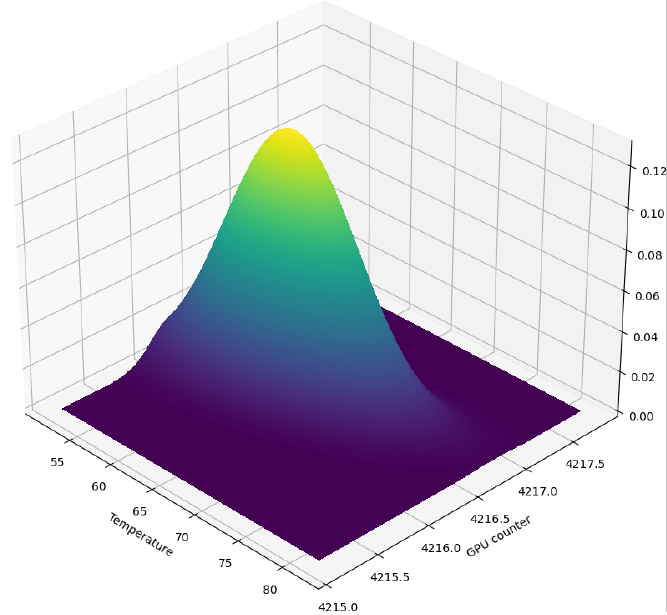}
    	\caption{Raspberry PI MAC dc:a6:32:4c:98:93}
	\end{subfigure}
	    \begin{subfigure}{0.45\textwidth}
    \includegraphics[width=\textwidth,trim={0 0 0 0} ,clip=true]{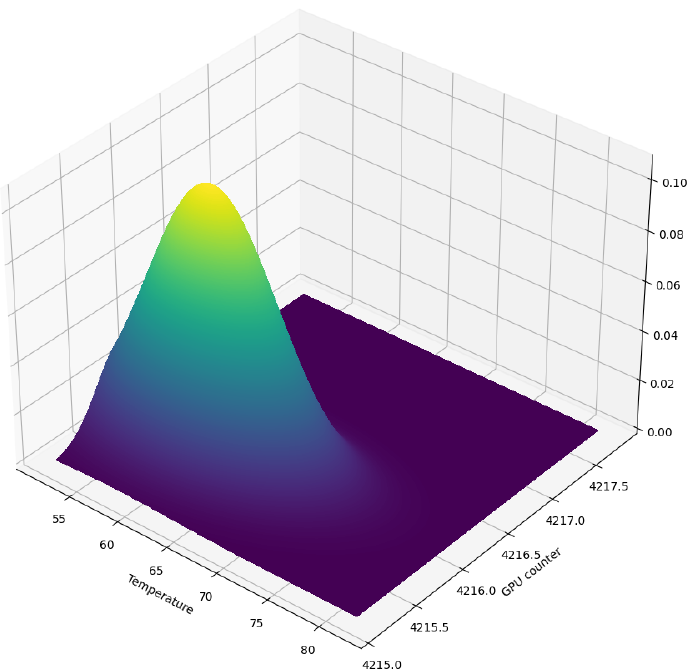}
    	\caption{Raspberry PI MAC dc:a6:32:4c:99:bf}
	\end{subfigure}
    \caption{3D density plot of temperature/GPU cycle counter value
for different devices executing the 120s-sleep function.}
    \label{fig:temp}
\end{figure*}

From these results, it can be concluded that the stability of these approaches is not sufficient for dynamic IoT scenarios where the devices operate in a typical way ( i.e. devices are restarted from time to time and power can go out). In contrast, they would be useful in IoT environments where device reboots are not possible, such as in the control of electrical or security systems.

\subsection{Lessons Learned and Limitations}

From the above comparison and the tests performed, valuable conclusions are drawn, both in the form of lessons learned and possible limitations of the proposed methodology. Regarding lessons learned, the main ones are:

\textbf{Component isolation is critical}. As \tablename~\ref{tab:validation} shows, it can be seen that isolating the measurements from external processes is crucial to ensure the \textit{stability} of the fingerprinting process. In this sense, in cases where the conditions of the components were not stable, it was not possible to reliably identify them after device rebooting.

\textbf{Rebooting can have impact on the fingerprints.} During the testing of literature-based validation approaches (see \tablename~\ref{tab:validation}), it was observed that the restart of the devices has an impact when the fingerprinting program is not isolated from other processes, probably due to the effect of the process scheduler. In contrast, this issue was not present in the approach of Section~\ref{sec:PoC}, as the data collection process was properly isolated from the noise introduced by other processes in the device. From this validation, it an be concluded that the \textit{robustness} property against the negative effects of other processes running in the device is achieved.

\textbf{Temperature does not seem to affect the components selected for validation}. The above tests have been performed at different temperature conditions and this condition does not seem to affect the results, possibly because by using integrated components on the same chip, it affects the base frequency and overall performance equally. Therefore, the \textit{robustness} property is met based on the temperature context. Actually, temperature was also measured during the data gathering of Section \ref{sec:PoC}. Using it as a feature, the average TPR for XGBoost is increased from 91.92\% to 93.46\%. Therefore, this information can be added as a correlation feature, incorporating supplementary information to the identification process. For different devices, \figurename~\ref{fig:temp} shows the density plot of the correlation in different devices of the temperatures and the GPU counter value after a 120-second CPU sleep. It can be seen that each device has a different plot shape and temperature is not influencing that the devices generate a similar fingerprint.

In terms of limitations of the methodology, the following have been identified after its design and validation:

\textbf{The methodology implementation is highly dependent on the hardware model}. The implementation of the present methodology, being based on the hardware components available in the devices, is highly dependent on the libraries needed to interact with them. Thus, implementations of the methodology may not be compatible between different models of single-board devices if their components are different, so it would be necessary to adapt the code.

\textbf{Some steps might need an exploratory analysis}. It is difficult to determine which hardware behavior measurements to take or which features to extract a priory. So, the implementation of the methodology may require several exploratory iterations to find a combination that meets all the properties needed in the generated fingerprint. This trial-and-error analysis can be highly reduced by analyzing the leveraged devices properties, different component and running frequencies. As every chip has imperfections, the challenge is how to measure them properly. In Section \ref{sec:PoC}, a successful application of the methodology has been provided, which serves as a guide and recommendation for future applications.

\textbf{Scalability in large deployments.} Manufacturing errors and variations are within the accepted tolerance range accepted by the manufacturers. Therefore, using these variations for identification in large deployments makes a single source of data possibly not sufficient \cite{polcak2015clock}. Thus, depending on the number of devices to be individually identified, a greater number of components and features should be employed to generate unique device fingerprints. Therefore, \textit{scalability} property arises as one of the most difficult properties to be met.

\section{CONCLUSIONS AND FUTURE WORK}
\label{sec:conclusions}

This paper proposes a methodology composed of seven steps that allow identifying identical single-board devices (same hardware and software configuration) used in heterogeneous IoT scenarios. These seven steps are grouped into two main phases, one to generate a behavioral fingerprint and another to evaluate it and identify the device. This work also presents the threat model affecting single-board device identification and seven properties that solutions dealing with identical device identification based on behavioral fingerprint must consider: uniqueness, stability, diversity, scalability, efficiency, robustness and security. The proposed methodology has been successfully validated in a real environment composed of 25 identical Raspberry Pi 4 Model B and Raspberry Pi 3 Model B+ using ML techniques for data processing. These devices were perfectly identified using a XGBoost model trained using features derived from the variation in performance between their CPU and GPU by setting a 50\% TPR threshold. Besides, this work compared the methodology identification performance with other implementations inspired in the literature works and provided some lessons learned and limitations.

As future work, it is planned to validate the methodology in larger scenarios with more devices and types, defining new features to be obtained and other ML/DL algorithms to evaluate the scalability of the solution in larger environment. Furthermore, it is desired to explore the usage of TEEs when generating the fingerprint, guaranteeing the security of the measurements by isolating the fingerprinting program from the rest of the system processes. Besides, we also plan to perform adversarial attacks against the proposed validation PoC, improving its resilience and performance.



\section*{ACKNOWLEDGMENT}


This work has been partially supported by \textit{(a)} the Swiss Federal Office for Defense Procurement (armasuisse) with the TREASURE and CyberSpec (CYD-C-2020003)  projects and \textit{(b)} the University of Zürich UZH.


\bibliographystyle{unsrt}
\bibliography{references}

\begin{thebibliography}{10}

\bibitem{fayos2020performance}
Rafael Fayos-Jordan, Santiago Felici-Castell, Jaume Segura-Garcia, Jesus
  Lopez-Ballester, and Maximo Cobos.
\newblock Performance comparison of container orchestration platforms with low
  cost devices in the fog, assisting internet of things applications.
\newblock {\em Journal of Network and Computer Applications}, 169:102788, 2020.

\bibitem{gomez2019generation}
A.~L. Perales~G{\'o}mez, L.~Fern{\'a}ndez~Maim{\'o}, A.~Huertas~Celdran, F.~J.
  Garc{\'\i}a~Clemente, C.~Cadenas~Sarmiento, C.~J. Del Canto~Masa, and
  R.~M{\'e}ndez~Nistal.
\newblock On the generation of anomaly detection datasets in industrial control
  systems.
\newblock {\em IEEE Access}, 7:177460--177473, 2019.

\bibitem{jagdale2022new}
Saumitra Jagdale.
\newblock {The Role of Hardware Root of Trust in Edge Devices}.
\newblock
  \url{https://www.eetimes.eu/the-role-of-hardware-root-of-trust-in-edge-devices/},
  2022.
\newblock [Online; accessed 21-June-2022].

\bibitem{Montalbano2022new}
Elizabeth Montalbano.
\newblock {Bluetooth Spoofing Bug Affects Billions of IoT Devices}.
\newblock
  \url{https://threatpost.com/bluetooth-spoofing-bug-iot-devices/159291/},
  2020.
\newblock [Online; accessed 21-June-2022].

\bibitem{liu2020zero}
Yongxin Liu, Jian Wang, Jianqiang Li, Houbing Song, Thomas Yang, Shuteng Niu,
  and Zhong Ming.
\newblock Zero-bias deep learning for accurate identification of
  internet-of-things (iot) devices.
\newblock {\em IEEE Internet of Things Journal}, 8(4):2627--2634, 2020.

\bibitem{nosouhi2022towards}
Mohammad~Reza Nosouhi, Keshav Sood, Marthie Grobler, and Robin Doss.
\newblock Towards spoofing resistant next generation iot networks.
\newblock {\em IEEE Transactions on Information Forensics and Security}, 2022.

\bibitem{yousefnezhad2020security}
Narges Yousefnezhad, Avleen Malhi, and Kary Fr{\"a}mling.
\newblock Security in product lifecycle of iot devices: A survey.
\newblock {\em Journal of Network and Computer Applications}, 171:102779, 2020.

\bibitem{sabhanayagam2022comparative}
T~Sabhanayagam.
\newblock A comparative analysis to obtain unique device fingerprinting.
\newblock In {\em Proceedings of International Conference on Deep Learning,
  Computing and Intelligence}, pages 349--354. Springer, 2022.

\bibitem{al2018survey}
Alauddin Al-Omary, Ali Othman, Haider~M AlSabbagh, and Hussain Al-Rizzo.
\newblock Survey of hardware-based security support for iot/cps systems.
\newblock {\em KnE Engineering}, pages 52--70, 2018.

\bibitem{polcak2015clock}
L.~Polc{\'a}k and B.~Frankov{\'a}.
\newblock Clock-skew-based computer identification: Traps and pitfalls.
\newblock {\em Journal of Universal Computer Science}, 21(9):1210--1233, 2015.

\bibitem{Sanchez2018clock}
I.~Sanchez-Rola, I.~Santos, and D.~Balzarotti.
\newblock Clock around the clock: Time-based device fingerprinting.
\newblock In {\em 2018 ACM SIGSAC Conference on Computer and Communications
  Security}, page 1502–1514, 2018.

\bibitem{sanchez2020survey}
P.~M. S\'anchez~S\'anchez, J.~M. Jorquera~Valero, A.~Huertas~Celdr\'an,
  G.~Bovet, M.~Gil~P\'erez, and G.~Mart\'inez~P\'erez.
\newblock A survey on device behavior fingerprinting: Data sources, techniques,
  application scenarios, and datasets.
\newblock {\em IEEE Communications Surveys \& Tutorials}, 23(2):1048--1077,
  2021.

\bibitem{code}
P.~M. S\'anchez~S\'anchez.
\newblock identification\_methodology.
\newblock \url{https://github.com/sxz0/identification\_methodology}, 2021.
\newblock [Online; accessed 8-June-2021].

\bibitem{salo2007multi}
T.~J. Salo.
\newblock Multi-factor fingerprints for personal computer hardware.
\newblock In {\em MILCOM 2007-IEEE Military Communications Conference}, pages
  1--7, October 2007.

\bibitem{jana2009skew}
S.~Jana and S.~K. Kasera.
\newblock On fast and accurate detection of unauthorized wireless access points
  using clock skews.
\newblock {\em IEEE Transactions on Mobile Computing}, 9(3):449--462, 2009.

\bibitem{sharma2012skew}
S.~Sharma, A.~Hussain, and H.~Saran.
\newblock Experience with heterogenous clock-skew based device fingerprinting.
\newblock In {\em 2012 Workshop on Learning from Authoritative Security
  Experiment Results}, pages 9--18, July 2012.

\bibitem{wang2012flash}
Y.~Wang, W.~Yu, S.~Wu, G.~Malysa, G.~E. Suh, and E.~C. Kan.
\newblock Flash memory for ubiquitous hardware security functions: True random
  number generation and device fingerprints.
\newblock In {\em 2012 IEEE Symposium on Security and Privacy}, pages 33--47,
  2012.

\bibitem{radhakrishnan2014gtid}
S.~V. Radhakrishnan, A.~S. Uluagac, and R.~Beyah.
\newblock {GTID}: A technique for physical device and device type
  fingerprinting.
\newblock {\em IEEE Transactions on Dependable and Secure Computing},
  12(5):519--532, 2014.

\bibitem{nakibly2015hardware}
G.~Nakibly, G.~Shelef, and S.~Yudilevich.
\newblock Hardware fingerprinting using {HTML5}.
\newblock {\em arXiv preprint arXiv:1503.01408}, 2015.

\bibitem{Jafari_Fingerprinting_Deep_Learning_2018}
H.~{Jafari}, O.~{Omotere}, D.~{Adesina}, H.~{Wu}, and L.~{Qian}.
\newblock {IoT} devices fingerprinting using deep learning.
\newblock In {\em 2018 IEEE Military Communications Conference}, pages 1--9,
  October 2018.

\bibitem{Riyaz2018Radio}
S.~{Riyaz}, K.~{Sankhe}, S.~{Ioannidis}, and K.~{Chowdhury}.
\newblock Deep learning convolutional neural networks for radio identification.
\newblock {\em IEEE Communications Magazine}, 56(9):146--152, 2018.

\bibitem{dong2019cpg}
S.~Dong, F.~Farha, S.~Cui, J.~Ma, and H.~Ning.
\newblock {CPG-FS}: A {CPU} performance graph based device fingerprint scheme
  for devices identification and authentication.
\newblock In {\em 4th IEEE Cyber Science and Technology Congress}, pages
  266--270, August 2019.

\bibitem{BabunCPS2021}
L.~Babun, H.~Aksu, and A.~S. Uluagac.
\newblock {CPS} device-class identification via behavioral fingerprinting: From
  theory to practice.
\newblock {\em IEEE Transactions on Information Forensics and Security},
  16:2413--2428, 2021.

\bibitem{babaei2019physical}
A.~Babaei and G.~Schiele.
\newblock Physical unclonable functions in the internet of things: State of the
  art and open challenges.
\newblock {\em Sensors}, 19(14):3208, 2019.

\bibitem{kong2013processor}
Joonho Kong and Farinaz Koushanfar.
\newblock Processor-based strong physical unclonable functions with aging-based
  response tuning.
\newblock {\em IEEE Transactions on Emerging Topics in Computing}, 2(1):16--29,
  2013.

\bibitem{gao2019building}
Yansong Gao, Yang Su, Wei Yang, Shiping Chen, Surya Nepal, and Damith~C
  Ranasinghe.
\newblock Building secure sram puf key generators on resource constrained
  devices.
\newblock In {\em 2019 IEEE International Conference on Pervasive Computing and
  Communications Workshops (PerCom Workshops)}, pages 912--917. IEEE, 2019.

\bibitem{kohno2005remote}
T.~Kohno, A.~Broido, and K.~C. Claffy.
\newblock Remote physical device fingerprinting.
\newblock {\em IEEE Transactions on Dependable and Secure Computing},
  2(2):93--108, 2005.

\bibitem{lanze2012skew}
F.~Lanze, A.~Panchenko, B.~Braatz, and A.~Zinnen.
\newblock Clock skew based remote device fingerprinting demystified.
\newblock In {\em 2012 IEEE Global Communications Conference}, pages 813--819,
  December 2012.

\bibitem{huang2014blueid}
J.~Huang, W.~Albazrqaoe, and G.~Xing.
\newblock {BlueID}: A practical system for {Bluetooth} device identification.
\newblock In {\em IEEE INFOCOM 2014-IEEE Conference on Computer
  Communications}, pages 2849--2857. IEEE, 2014.

\bibitem{pawar2017wide}
S.~N. Pawar and P.~B. Mane.
\newblock Wide band {PLL} frequency synthesizer: A survey.
\newblock In {\em 2017 International Conference on Advances in Computing,
  Communication and Control}, pages 1--6. IEEE, 2017.

\bibitem{marabissi2022iot}
Dania Marabissi, Lorenzo Mucchi, and Andrea Stomaci.
\newblock Iot nodes authentication and id spoofing detection based on joint use
  of physical layer security and machine learning.
\newblock {\em Future Internet}, 14(2):61, 2022.

\bibitem{rajan2017sybil}
Anjana Rajan, J~Jithish, and Sriram Sankaran.
\newblock Sybil attack in iot: Modelling and defenses.
\newblock In {\em 2017 international conference on advances in computing,
  communications and informatics (ICACCI)}, pages 2323--2327. IEEE, 2017.

\bibitem{chen2022machine}
Zhiyan Chen, Jinxin Liu, Yu~Shen, Murat Simsek, Burak Kantarci, Hussein~T
  Mouftah, and Petar Djukic.
\newblock Machine learning-enabled iot security: Open issues and challenges
  under advanced persistent threats.
\newblock {\em ACM Computing Surveys (CSUR)}, 2022.

\bibitem{li2020adversarial}
Deqiang Li and Qianmu Li.
\newblock Adversarial deep ensemble: Evasion attacks and defenses for malware
  detection.
\newblock {\em IEEE Transactions on Information Forensics and Security},
  15:3886--3900, 2020.

\bibitem{ruhrmair2012security}
Ulrich R{\"u}hrmair, Srinivas Devadas, and Farinaz Koushanfar.
\newblock Security based on physical unclonability and disorder.
\newblock In {\em Introduction to Hardware Security and Trust}, pages 65--102.
  Springer, 2012.

\bibitem{sembiring2021randomness}
Rivaldo~Ludovicus Sembiring, Rizka~Reza Pahlevi, and Parman Sukarno.
\newblock Randomness, uniqueness, and steadiness evaluation of physical
  unclonable functions.
\newblock In {\em 2021 9th International Conference on Information and
  Communication Technology (ICoICT)}, pages 429--433. IEEE, 2021.

\bibitem{hamza2018clear}
Ayyoob Hamza, Dinesha Ranathunga, Hassan~Habibi Gharakheili, Matthew Roughan,
  and Vijay Sivaraman.
\newblock Clear as mud: Generating, validating and applying iot behavioral
  profiles.
\newblock In {\em Proceedings of the 2018 Workshop on IoT Security and
  Privacy}, pages 8--14, 2018.

\bibitem{ahmed2022analyzing}
Dilawer Ahmed, Anupam Das, and Fareed Zaffar.
\newblock Analyzing the feasibility and generalizability of fingerprinting
  internet of things devices.
\newblock {\em Proceedings on Privacy Enhancing Technologies},
  2022(2):578--600, 2022.

\bibitem{arellanes2020evaluating}
Damian Arellanes and Kung-Kiu Lau.
\newblock Evaluating iot service composition mechanisms for the scalability of
  iot systems.
\newblock {\em Future Generation Computer Systems}, 108:827--848, 2020.

\bibitem{peng2018toward}
Limei Peng, Ahmad~R Dhaini, and Pin-Han Ho.
\newblock Toward integrated cloud--fog networks for efficient iot provisioning:
  Key challenges and solutions.
\newblock {\em Future Generation Computer Systems}, 88:606--613, 2018.

\bibitem{zhou2019design}
Xinyu Zhou, Aiqun Hu, Guyue Li, Linning Peng, Yuexiu Xing, and Jiabao Yu.
\newblock Design of a robust rf fingerprint generation and classification
  scheme for practical device identification.
\newblock In {\em 2019 IEEE Conference on Communications and Network Security
  (CNS)}, pages 196--204. IEEE, 2019.

\bibitem{lu2018internet}
Yang Lu and Li~Da~Xu.
\newblock Internet of things (iot) cybersecurity research: A review of current
  research topics.
\newblock {\em IEEE Internet of Things Journal}, 6(2):2103--2115, 2018.

\bibitem{arp2022and}
Daniel Arp, Erwin Quiring, Feargus Pendlebury, Alexander Warnecke, Fabio
  Pierazzi, Christian Wressnegger, Lorenzo Cavallaro, and Konrad Rieck.
\newblock Dos and don’ts of machine learning in computer security.
\newblock In {\em Proc. of the USENIX Security Symposium}, 2022.

\bibitem{harris2010transactional}
T.~Harris, J.~Larus, and R.~Rajwar.
\newblock Transactional memory.
\newblock {\em Synthesis Lectures on Computer Architecture}, 5(1):1--263, 2010.

\bibitem{lee2020softee}
U.~Lee and C.~Park.
\newblock {SofTEE}: Software-based trusted execution environment for user
  applications.
\newblock {\em IEEE Access}, 8:121874--121888, 2020.

\bibitem{cadavid2020machine}
J.~P. Usuga~Cadavid, S.~Lamouri, B.~Grabot, R.~Pellerin, and A.~Fortin.
\newblock Machine learning applied in production planning and control: A
  state-of-the-art in the era of industry 4.0.
\newblock {\em Journal of Intelligent Manufacturing}, pages 1--28, 2020.

\bibitem{undocumentedPi}
Embedded~Linux Wiki.
\newblock The undocumented pi.
\newblock \url{https://elinux.org/The_Undocumented_Pi/}, 2021.
\newblock [Online; accessed 8-June-2021].

\bibitem{py-videocore6}
Idein.
\newblock py-videocore6. {Python} library for {GPU} programming on {Raspberry
  Pi} 4.
\newblock \url{https://github.com/Idein/py-videocore6/}, 2021.
\newblock [Online; accessed 8-June-2021].

\bibitem{py-videocore}
Idein.
\newblock py-videocore. {Python} library for {GPGPU} on {Raspberry Pi}.
\newblock \url{https://github.com/nineties/py-videocore/}, 2021.
\newblock [Online; accessed 8-June-2021].

\bibitem{op-tee}
TrustedFirmware.org.
\newblock {OP-TEE documentation. Raspberry Pi 3}.
\newblock
  \url{https://optee.readthedocs.io/en/latest/building/devices/rpi3.html/},
  2012.
\newblock [Online; accessed 8-June-2022].

\end{thebibliography}

\end{document}